\documentclass[useAMS,usenatbib]{mnras}
\usepackage[a4paper]{geometry}
\usepackage{fixltx2e}
\usepackage{graphicx}
\usepackage{rotating}
\usepackage{pdflscape}
\usepackage[T1]{fontenc}
\usepackage{ae,aecompl}
\usepackage{hyperref}

\usepackage{color}
\usepackage{amsmath}
\usepackage{amssymb}
\usepackage[update]{epstopdf}
\title[Cyclotron Maser Instability at Hot Jupiters]{How Expanded Ionospheres of Hot Jupiters Can Prevent Escape of Radio Emission Generated by the Cyclotron Maser Instability}
\author[C. Weber et al.]
{C.~Weber$^{1}$, H.~Lammer$^1$, I. F.~Shaikhislamov$^2$, J. M.~Chadney$^{3,4}$,
\newauthor M. L.~Khodachenko$^1$, J.-M.~Grie\ss meier$^5$, H. O.~Rucker$^6$, C.~Vocks$^7$,
\newauthor  W. Macher$^1$, P.~Odert$^1$, K. G.~Kislyakova$^{8,1}$
\\
$^1$Space Research Institute, Austrian Academy of Sciences, Schmiedlstr. 6, A-8042, Graz, Austria\\
$^2$Institute of Laser Physics SB RAS, Novosibirsk, Russia\\
$^3$Department of Physics and Astronomy, University of Southampton, Southampton SO17 1BJ, UK\\
$^4$Department of Physics, Imperial College London, Prince Consort Road, London SW7 2BW, UK\\
$^5$LPC2E - Universit\'{e} d'Orl\'{e}ans/CNRS, France and Station de Radioastronomie de Nan\c{c}ay, Observatoire de Paris, \\ PSL Research University, CNRS/INSU, USR 704 - Univ. Orl\'{e}ans, OSUC, route de Souesmes, 18330 Nan\c{c}ay, France\\
$^6$Comm. Astron. \"OAW, Graz, Austria\\
$^7$Leibniz-Inst. f. Astrophys. Potsdam, Germany\\
$^8$University of Vienna, Department of Astrophysics, T\"{u}rkenschanzstrasse 17, A-1180 Vienna, Austria}

\date{Accepted XXX. Received YYY; in original form ZZZ}

\pubyear{2017}

\begin{document}
\label{firstpage}
\pagerange{\pageref{firstpage}--\pageref{lastpage}}
\maketitle

\begin{abstract}
We present a study of plasma conditions in the atmospheres of the Hot Jupiters HD 209458b and HD 189733b and for an HD 209458b-like planet at orbit locations between 0.2--1 AU around a Sun-like star. We discuss how these conditions influence the radio emission we expect from their magnetospheres. We find that the environmental conditions are such that the cyclotron maser instability (CMI), the process responsible for the generation of radio waves at magnetic planets in the solar system, most likely will not operate at Hot Jupiters. Hydrodynamically expanding atmospheres possess extended ionospheres whose plasma densities within the magnetosphere are so large that the plasma frequency is much higher than the cyclotron frequency, which contradicts the condition for the production of radio emission and prevents the escape of radio waves from close-in exoplanets at distances $<$0.05 AU from a Sun-like host star. The upper atmosphere structure of gas giants around stars similar to the Sun changes between 0.2 and 0.5 AU from the hydrodynamic to a hydrostatic regime and this results in conditions similar to solar system planets with a region of depleted plasma between the exobase and the magnetopause where the plasma frequency can be lower than the cyclotron frequency. In such an environment, a beam of highly energetic electrons accelerated along the field lines towards the planet can produce radio emission. However, even if the CMI could operate the extended ionospheres of Hot Jupiters are too dense to let the radio emission escape from the planets.
\end{abstract}


\section{Introduction}

Already in the late seventies and in the eighties, before the first exoplanet had been discovered, the search for radio emission from exoplanets had started with observations by \citet{Yantis1977} and  \citet{Winglee1986}. Various attempts of detection have been performed for known exoplanets \citep{Zarka1997,Ryabov2004,Zarka2011,Bastian2000,Farrell2003,Farrell2004,Majid2006,Winterhalter2006,Griessmeier2006b,George2007,Smith2009,Lecavelier2009,Lecavelierdesetangs2011,Lecavelierdesetangs2013,Lazio2007,Lazio2010,Hallinan2012,Griessmeier2017,Turner2017}. \citet{Sirothia2014} used archival survey data from the GMRT (Giant Metrewave Radio Telescope) to search for exoplanets in the radio wavelength range. So far, this search is still unsuccessful. \citet{Lecavelierdesetangs2013} found hints of radio emission from the Neptune-mass extrasolar planet HAT-P-11b at 150 MHz and \citet{Sirothia2014} found hints of emission with the GMRT of the Tata Institute of Fundamental Research (TIFR GMRT) Sky Survey from various objects but these possible detections remain to be confirmed.

Many authors presented predictions of radio fluxes and radio powers for exoplanets based on scaling laws for estimated magnetic fields of the respective planets \citep{Farrell1999,Farrell2004,Zarka1997,Zarka2001,Zarka2004,Zarka2007,Lazio2004,Griessmeier2004,Griessmeier2005,Griessmeier2007c,Griessmeier2007b,Griessmeier2011,Stevens2005,Jardine2008,Fares2010,Reiners2010, Nichols2011,Nichols2012,Vidotto2012,See2015b,See2015a}. However, the exoplanetary magnetic fields are still unconstrained by observations. This might change via the detection of radio emission from exoplanets because the direct relation of the measured frequency of this emission with the magnetic field strength of the planet makes the indirect detection of exoplanetary magnetic fields possible \citep{Griessmeier2015}.

The aforementioned estimations of exoplanetary radio emission did usually not consider propagation effects related to exoplanetary ionospheres. Only the surrounding plasma environment of the planet, i.e. the stellar wind, was considered as an obstacle for escape of the radio waves from the source region \citep{Griessmeier2007b}. \citet{Koskinen2013a} discuss briefly the propagation and find that, if the emission is generated in the ionosphere at 1 to 5 planetary radii, emission below 10 to 70 MHz is blocked by the planetary ionosphere. \citet{Fujii2016} consider possible radio emission from Hot Jupiters around Red Giant stars. For the implications for detectability they consider the plasma frequency of the stellar wind and of the Earth's ionosphere as a cutoff frequency for the radio emission and refer to a future study concerning the plasma frequency or density in a possible source region. \citet{Nichols2016} consider the interaction between stellar wind and Earth-like planetary magnetospheres, i.e. radio emission is resulting from magnetic reconnection processes in the tail. In their work, detailed ionosphere models for strongly irradiated Hot Jupiters are not taken into account. The main aim of our study is to include the results obtained from ionosphere models for close-in extrasolar gas giants and to check whether radio emission can escape from the source region or whether the Cyclotron Maser Instability (CMI) can occur at all. 

Over the past decade, several studies applied hydrodynamical upper atmosphere models which included photochemistry, ionization and dissociation to Hot Jupiters \citep{Yelle2004,Munoz2007,Koskinen2010,Koskinen2013a,Koskinen2013b,Koskinen2014,Koskinen2013,Guo2011,Shaikhislamov2014,Khodachenko2015,Salz2015,Chadney2015,Chadney2016,Erkaev2016}. In our study we use ionospheric profiles for Hot Jupiters modelled in previous studies of \citet{Shaikhislamov2014} and \citet{Khodachenko2015} for HD 209458b, and \citet{Guo2011} for HD 189733b. Moreover, we study the effect of orbit locations between 0.2 and 1 AU by using modelled plasma electron densities of \citet{Chadney2015, Chadney2016} for an HD 209458b-like planet at these locations. In Section \ref{sec:sec2}, we discuss the expected magnetic moments and corresponding magnetospheres of HD 209458b and HD 189733b, which is important for the calculation of the cyclotron frequencies in the exoplanet's magnetosphere. Then in Section \ref{sec:sec31} the basics of the Cyclotron Maser Instability are presented and the necessary conditions for generation and/or escape of radio waves are discussed. Sections \ref{sec:sec321} and \ref{sec:sec32} briefly describe the magnetic field and plasma environment model of \citet{Shaikhislamov2014} and \citet{Khodachenko2015} for HD 209458b. This leads directly to our results for the plasma and cyclotron frequencies for Hot Jupiters and an HD 209458b-like planet around a Sun-like star between 0.2--1 AU (Section \ref{sec:sec4}). Section \ref{sec:sec5} comprises a discussion and the conclusion of our work.

\section{Relevance of the Magnetic Field for Radio Emissions}
\label{sec:sec2}

Two of the most important advantages of detections of radio emission generated at exoplanets would be:

\begin{itemize}
\item Offering a way for detecting exoplanets directly;
\item Providing the most promising way to detect and quantify the magnetic field of the exoplanet \citep[e.g.][]{Griessmeier2015}.
\end{itemize}

Although exoplanets are discovered in large amount by other techniques like the radial velocity and transit method, radio observations would be a unique tool for the investigation of the magnetic moment and magnetospheres of planets outside the solar system.

 The relation between the maximum frequency of the emission and the maximum magnetic field strength is given by
\begin{equation}
	\label{eqn:freqbrelation}
	f_{\rm {c}} = \frac{1}{2\pi}\frac{e B}{m_{\rm e}}.
\end{equation}
Thus, from a radio measurement the magnetic field strength $B$ can be directly deduced. Here, $e$ is the electron charge and $m_{\rm e}$ is the electron mass.

The planetary magnetic moment is an ill-constrained, yet important quantity for estimating exoplanetary radio flux. Different theoretical arguments have led to two main approaches: \citet{Farrell1999} and \citet{Griessmeier2004} assume the planetary magnetic moment can be calculated by a force balance, and find a planetary magnetic field which depends on the planetary rotation rate. On the other hand, \citet{Reiners2010} assume the planetary magnetic moment to be primarily driven by the energy flux from the planetary core. Thus, they find no dependence on the planetary rotation rate; however, they obtain stronger magnetic fields and thus more favourable observing conditions for young planets. As a direct consequence, tidal locking has a strong influence on the planetary magnetic moment for the former model; for the latter, it is without consequence. For HD 209458b, \citet{Griessmeier2004} estimate the magnetic moment to be $\sim 0.1 \mathcal{M}_{\rm J}$ \citep[in agreement with the value estimated from the Ly-$\alpha$-HST observations by][]{Kislyakova2014}, where $\mathcal{M}_{\rm J} = 1.56 \cdot 10^{27}\textnormal{Am}^2$ is the Jovian magnetic moment. If the planet is not tidally locked, or if the magnetic moment does not depend on planetary rotation, the magnetic moment is estimated as $\sim 0.3 \mathcal{M}_{\rm J}$ \citep{Griessmeier2004}, or could be even higher according to \citet{Reiners2010}.

We can also calculate the maximum emitted frequency using Equation (\ref{eqn:freqbrelation}), and based on the estimations of the magnetic moment \citep{Griessmeier2004, Kislyakova2014}. These are $\mathcal{M} = 0.1 \mathcal{M}_{\rm J} = 1.6 \cdot 10^{26} \textnormal{Am}^2$ \citep{Kislyakova2014} for HD 209458b and $0.3 \mathcal{M}_{\rm J}$ \citep{Griessmeier2007b} for HD 189733b. Using
\begin{equation}
	\label{eqn:maxind}
	B = \frac{\mu_0}{4\pi}\frac{2 \mathcal{M}}{R_{\rm p}^3}
\end{equation}
\citep{Griessmeier2005,Griessmeier2007c,Griessmeier2007b} where $R_{\rm p}$ is the planetary radius (see Table \ref{tab:tab1}), and Equation (\ref{eqn:freqbrelation}) gives a maximum emission frequency of 0.9 MHz for HD 209458b and 4.9 MHz for HD 189733b, which is below the Earth's ionospheric cutoff of 10 MHz and not detectable from the ground. In the model by \citet{Reiners2010} the dipole field strength at the pole for HD 189733b is predicted to be 0.0014 T (14 G) (magnetic moment of $3.7 \cdot 10^{27}\textnormal{Am}^2 \approx 2.4 \mathcal{M}_{\rm J}$), i.e. observation of corresponding radio emission (39.19 MHz) would be possible. In Section \ref{sec:sec4} we compare the model with tidal locking and without tidal locking for HD 189733b. 

We also test a planet with a very strong magnetic moment, i.e. a young, massive planet (of age 100 Myr and with 13 Jupiter masses, i.e. on the boundary of being a brown dwarf) in the model where tidal locking has no influence on the magnetic field. According to \citet{Reiners2010} (their Figure 1), such a planet would have a magnetic moment of $\sim 50 \mathcal{M}_{\rm J}$ ($\sim 100 \mathcal{M}_{\rm J}$ for the same planet with 13 Jupiter masses at a very young age of 10 Myr). The radius of such a planet roughly equals the Jovian radius. The results are not shown in the plots but are discussed in Section \ref{sec:hm}.

Besides the uncertainty of the planetary magnetic moment a further point mentioned by \cite{Koskinen2013a} and investigated in detail here is that the radio emission, if it is produced at all, may have a problem to escape from the exoplanet because its own ionosphere would block the radiation.

\subsection{Magnetopause standoff distance}\label{sec:sec21}

If only the dipole approximation is considered then the magnetopause standoff distance $R_s$ is related to the magnetic moment via the formula \citep[e.g.][]{Griessmeier2004, Khodachenko2012, Kislyakova2014}
\begin{equation}
	\label{eqn:magmom}
	R_{\rm s} = \left(\frac{\mathcal{M}^2\mu_0 f_0^2}{8\pi^2\rho_{\rm{sw}}v_{\rm sw}^2}\right)^{1/6}.
\end{equation}
Here, $f_0 = 1.22$ is a form factor for the magnetopause shape including the influence of a magnetodisk \citep{Khodachenko2012}, $v_{\rm{sw}}$ is the relative velocity of the stellar wind corrected for the orbital motion of the planet, $\rho_{\rm{sw}}$ is the stellar wind density and $\mu_0$ is the vacuum permeability. The magnetopause standoff distance for HD 209458b is approximately $2.8 R_{\rm p}$ for the magnetic moment of $0.1 \mathcal{M}_{\rm J}$ estimated by \citet{Kislyakova2014}. For $\mathcal{M} = \mathcal{M}_{\rm J}$ it would be $\approx 6.1 R_{\rm p}$ for the same stellar wind conditions.

For HD 189733b with magnetic moment of $0.3 \mathcal{M}_{\rm J}$ an estimation of $R_{\rm s}$ from Equation (\ref{eqn:magmom}) gives $2.2 R_{\rm p}$ and for the Jovian magnetic moment $3.2 R_{\rm p}$ for the stellar wind parameters of Table \ref{tab:tab1}. These parameters stem from the stellar wind model used in \citet{Griessmeier2007c} where it was mentioned that for stars younger than 0.7 Gyr the results for stellar wind velocity and density are questionable. For a young, massive planet of $\sim 50 \mathcal{M}_{\rm J}$ (as mentioned above) we get $22.5 R_{\rm p}$ as standoff distance and $28.4 R_{\rm p}$ for $\sim 100 \mathcal{M}_{\rm J}$ for the same wind conditions as for HD 209458b.
\begin{table}
\caption{Parameters from the best fit of HST Ly-$\alpha$ observations for HD 209458b from \citet{Kislyakova2014} and of HD 189733b (orbital distance, mass and radius from \url{http://exoplanet.eu/catalog/hd_189733_b/}, accessed last time on 22.11.2016).}
\label{tab:tab1}
\begin{tabular}{p{3.5 cm}|p{2 cm}|p{2 cm}}
 & HD 209458b & HD 189733b \\
\hline
Orbital distance & $0.047$ AU & 0.03142 AU \\
Planetary Mass & $0.69 M_{\rm J}$ & $1.142 M_{\rm J}$ \\
Planetary Radius & $1.38 R_{\rm J}$ & $1.138 R_{\rm J}$ \\
Dipole moment $\mathcal{M}$ & $0.1 \mathcal{M}_{\rm J}$$^\ast$ & $0.3 \mathcal{M}_{\rm J}$$^{\ast\ast}$ \\
Radio frequency $f_{\rm{c}}$ & 1.032 MHz & 4.86 MHz \\
Standoff distance & 2.8 $R_{\rm p}$ & 2.15 $R_{\rm p}$\\
Stellar wind number density & $5 \cdot 10^9 \textnormal{m}^{-3}$ & $3.39 \cdot 10^{11} \textnormal{m}^{-3}$ \\
Stellar wind velocity & $426 \cdot 10^3$ m/s & $555 \cdot 10^3$ m/s\\
\hline
\end{tabular}
*estimated via Ly$\alpha$ measurement \citep{Kislyakova2014}
**estimated from scaling laws \citep{Griessmeier2007b}  
\end{table}

\citet{Khodachenko2012} provided an estimation for the standoff distance of a Hot Jupiter's magnetosphere including a magnetodisk. Using the parameters of HD 209458b and HD 189733b as given in Table \ref{tab:tab1} and the same stellar wind parameters as above we can calculate the standoff distance for this case using the formula given in \citet{Khodachenko2012}:
\begin{align}
\nonumber
\frac{R_{\rm s}}{R_{\rm p}} \sim & \frac{B_{\rm{d0J}}^{1/2}\left(1 + \kappa^2\right)^{1/4}}{\left(2\mu_0 p_{\rm{sw}}\right)^{1/4}}\left(\frac{R_{\rm{AJ}}}{R_{\rm p}}\right)^{-1/2} \times \\
\label{eqn:standoffmd}
& \times \left(\frac{\omega_{\rm{p}}}{\omega_{\rm J}}\right)^{(3 k + 1)/10}\left(\frac{dM_{\rm p}^{\left(th\right)}/dt}{dM_{\rm J}/dt}\right)^{1/10}.
\end{align}
The equatorial dipole field strength of Jupiter is $B_{\rm{d0J}} = 0.000428$ T, $\kappa = 2f_0$, $p_{\rm{sw}}$ is the solar wind ram pressure and is given by $v_{\rm{sw}}^2 \cdot \rho_{\rm{sw}}$, where $\rho_{\rm{sw}}$ is the stellar wind density and $v_{\rm{sw}}$ the velocity. $R_{\rm{AJ}} = 19.8 R_{\rm J}$ is the Alfv\'{e}nic radius of Jupiter. The Jovian and the planetary angular velocity of rotation are given by $\omega_{\rm J}$ and $\omega_{\rm p}$ and $dM_{\rm p}^{\left(th\right)}/dt = 1.06 \cdot 10^{7} \textnormal{kg/s}$ and $dM_{\rm J}/dt = 10^3 \textnormal{kg/s}$ are the thermal mass loss rate of the planet and the mass loss rate of Jupiter, respectively \citep{Khodachenko2012}. The power index $k$ is $1/2$. This yields a standoff distance of about $8.2 R_{\rm p}$ for HD 209458b, which is not in agreement with the best fit standoff distance given by \citet{Kislyakova2014} to be $\sim 2.8 R_{\rm p}$. Because we use different stellar wind parameters than \citet{Khodachenko2012} we also get a different standoff distance compared to their work. From this estimate one could conclude that HD 209458b might not have an additional relevant magnetic component caused by a magnetodisk. For HD 189733b we get a value of $3.8 R_{\rm p}$, slightly larger than the value without a magnetodisk. For this case the mass loss rate was taken from \citet{Guo2011} as $1.98 \cdot 10^{8} \textnormal{kg/s}$, valid for an UV flux of 100 Wm$^{-2}$, although it has to be mentioned that these authors do not consider a magnetized case.   

\section{Theoretical Background and Plasma Environment}
\label{sec:sec3}

To analyze the possibility of producing radio emission via the Cyclotron Maser Instability at Hot Jupiters and, if it operates, to check whether the emitted radio waves can escape from the source region, one needs first to determine the value of several important parameters. One of them is the plasma density in the source region because one of the conditions for the cyclotron maser to operate is that the plasma frequency should be lower than the cyclotron frequency, $f_{\rm{p}} \lesssim 0.4 f_{\rm{c}}$ \citep{Griessmeier2007b}, i.e. we need a plasma depleted region through which a beam of highly energetic electrons is propagating to produce radio emission (see Section \ref{sec:sec31}). Another important parameter is the strength of the planetary magnetic field. Without observations they are hard to constrain. This has been discussed in Section \ref{sec:sec2}.

Using plasma density profiles modeled by \citet{Shaikhislamov2014}, \citet{Khodachenko2015} and \citet{Guo2011} for the Hot Jupiters HD 209458b and HD 189733b, respectively and by \citet{Chadney2015, Chadney2016} for an HD 209458b-like planet at orbit locations between 0.2 and 1 AU together with magnetic moments estimated by \citet{Kislyakova2014} and \citet{Griessmeier2005,Griessmeier2007b} we can compare the plasma frequency and the cyclotron frequency and check whether the condition for production of radio emission via this process is fulfilled or not.

\subsection{The Cyclotron Maser Instability (CMI)}
\label{sec:sec31}

The basics of the radiation generation are the following: electrons from the solar or the stellar wind enter the magnetospheres in the cleft regions (regions of funnel shaped magnetic field structure) and are forced to follow the field lines on helical paths, i.e. they gyrate around the magnetic field lines at high latitudes \citep{Zarka1992,Zarka1998,Ray2008,Hess2008,Hess2010a,Hess2010b,Hess2011}. The sources of the radio emission are distributed along the high latitude magnetic field lines and emission is always close to the local gyrofrequency of the electrons (Equation \ref{eqn:freqbrelation}) \citep{Wu1979,Wu1985,Zarka1992,Zarka1998,Treumann2006}.

The component of electron movement perpendicular to $\bmath{B}$ (the gyration) is, if there is resonance with the electric field of a wave, in a constant phasing, which makes energy transfer from electrons to waves possible. The resonant interaction of the energetic electrons with a plasma wave in the magnetosphere leads to direct conversion of the electron's energy to electromagnetic wave energy via the so-called Cyclotron Maser Instability (CMI) mechanism. The CMI converts up to 1 \% of the free energy in the unstable electron distribution directly to electromagnetic waves \citep{Wu1979,Wu1985,Zarka1992,Zarka1998,Treumann2006}. 

The basic theory of the CMI was first studied by \citet{Wu1979}. Extensive reviews can be found in \citet{Wu1985} and \citet{Treumann2006}. The best conditions for the CMI are a background plasma with a very low density and a very strong background magnetic field, i.e.
\begin{equation}
	\label{eqn:densitylow}
	f_{\rm{p}} \ll f_{\rm{c}}.
\end{equation}
Since 
\begin{equation}
\label{eqn:fp}
f_{\rm{p}} = \frac{\omega_{\rm{p}}}{2\pi} = \frac{1}{2\pi}\sqrt{\frac{e^2n_{\rm e}}{m_{\rm e}\varepsilon_0}}, 
\end{equation}
where $n_{\rm e}$ is the electron density, $m_{\rm e}$ the electron mass and $\varepsilon_0$ the vacuum permittivity, we see that the plasma frequency is directly connected with the plasma density which tells us that condition (\ref{eqn:densitylow}) means that the plasma should be dilute. It should be noted that the CMI also operates for $f_{\rm{p}} \leq f_{\rm{c}}$ but then it is less efficient than for $f_{\rm{p}} \ll f_{\rm{c}}$ \citep{Treumann2006}. The largest possible ratio is $f_{\rm{p}}/f_{\rm{c}} = 0.4$ \citep{Lequeau1983,Lequeau1985,Treumann2006,Griessmeier2007b}. The generation of radio radiation also requires the existence of an electric field component $\bmath{E}$ parallel to the magnetic field $\bmath{B}$ \citep{Treumann2006} to accelerate the electrons along the magnetic field lines towards the planet. In general the CMI operates in plasmas with $f_{\rm{p}}/f_{\rm{c}} \leq 0.4$. In Section \ref{sec:sec32} and Section \ref{sec:sec4} we will check whether the plasma environment provides the necessary conditions for the CMI to operate.

All planets in the solar system with a strong magnetic field are sources of intense radio emission with origin along magnetic field lines in high latitudes. So the basic ingredients for the planets to be radio emitters are their dynamo action, which leads to the generation of a magnetic field and thus to the development of an extended magnetosphere for the five radio planets Earth, Jupiter, Saturn, Uranus and Neptune, and the input of energetic electrons into the magnetosphere. As was already mentioned this happens primarily via the solar wind but for Jupiter the Galilean moon Io acts also as an internal plasma source, leading to strong DAM (decametric) radio emission up to 40 MHz. These emissions all have a high degree of circular polarization, in most cases up to 100 \%, and the main emission mode is the X-mode (right-handed extraordinary mode) \citep{Treumann2006}.

For close-in giant exoplanets it is believed that the radio emission is mainly triggered by the stellar wind \citep{Zarka2001}. The wind's particles enter the magnetosphere in the nightside cusp region and are transported into the radio source regions via reconnection processes between the exoplanetary and the stellar field lines.

\subsection{Density and Magnetic Field Model}
\label{sec:sec321}

The evaluated data presented in Section \ref{sec:sec4} for HD 209458b's densities and magnetic field comes from the model by \citet{Khodachenko2015}. It extends the model from \citet{Shaikhislamov2014}, which did not consider magnetic fields, to describe structures of the inner magnetosphere of a Hot Jupiter, which is formed by the planetary plasma wind and the planetary dipole magnetic field. Additionally to the effect from the magnetic field the model by \citet{Khodachenko2015} includes a realistic spectrum of the stellar XUV radiation, basic hydrogen chemistry, H$_3^+$-cooling and tidal as well as rotational forces between the star and the planet.

The model equations which are solved in \citet{Khodachenko2015} are the same as in \citet{Shaikhislamov2014} but also include equations for hydrogen chemistry and equations for the magnetic field as well as extended momentum and energy equations. The latter include Ampere forces and general radiation heating terms.

The magnetic field is poloidal. It is calculated in a planetary centered cylindrical coordinate system by propagating an axisymmetric vector potential in time. The $z$-axis of this coordinate system points along the dipole moment. For further details on the model we refer to \citet{Shaikhislamov2014} and \citet{Khodachenko2015}.    

\subsection{Plasma environment}
\label{sec:sec32}

\begin{figure*}
\begin{center}
\includegraphics[width=0.6\columnwidth]{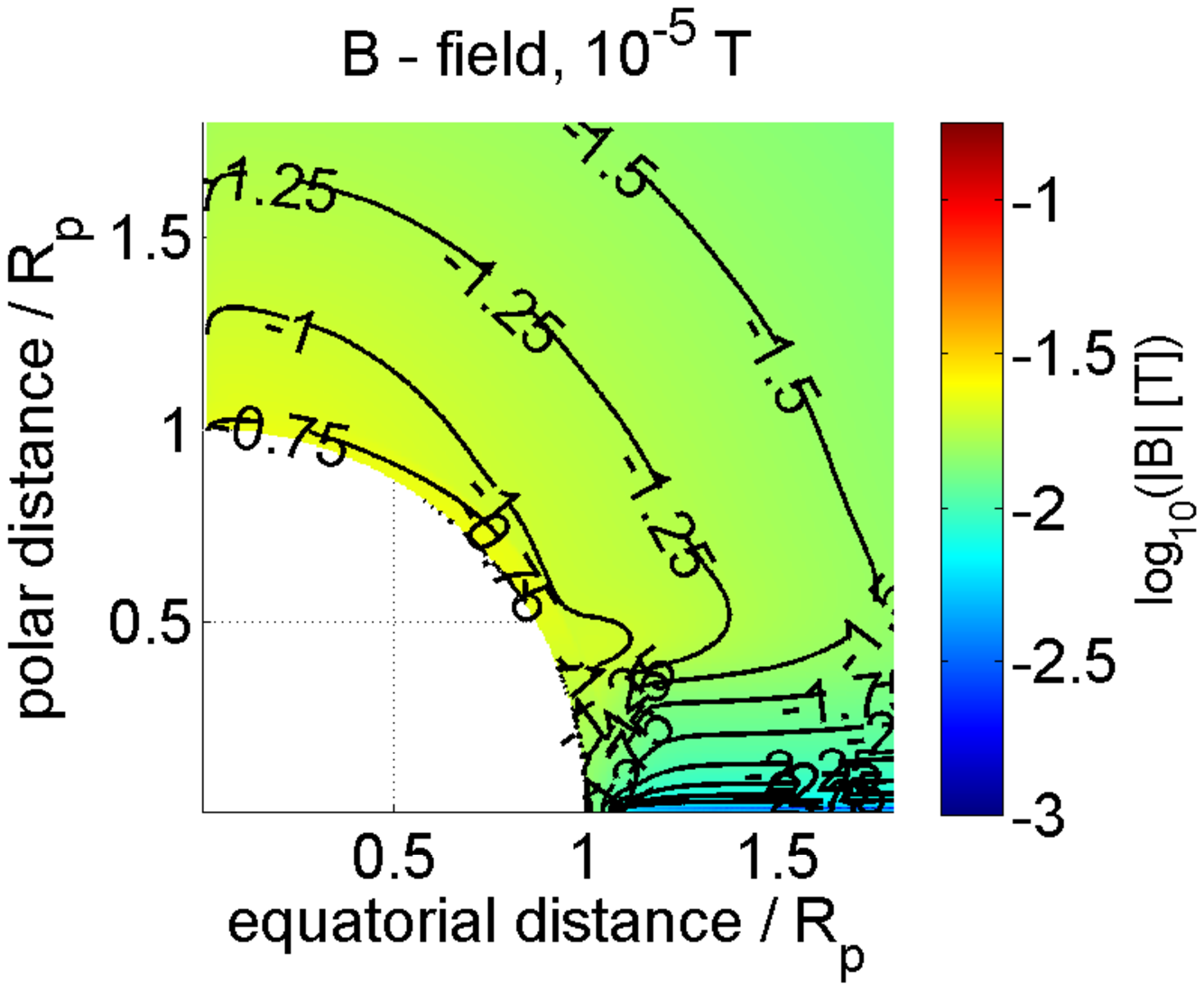}
\includegraphics[width=0.6\columnwidth]{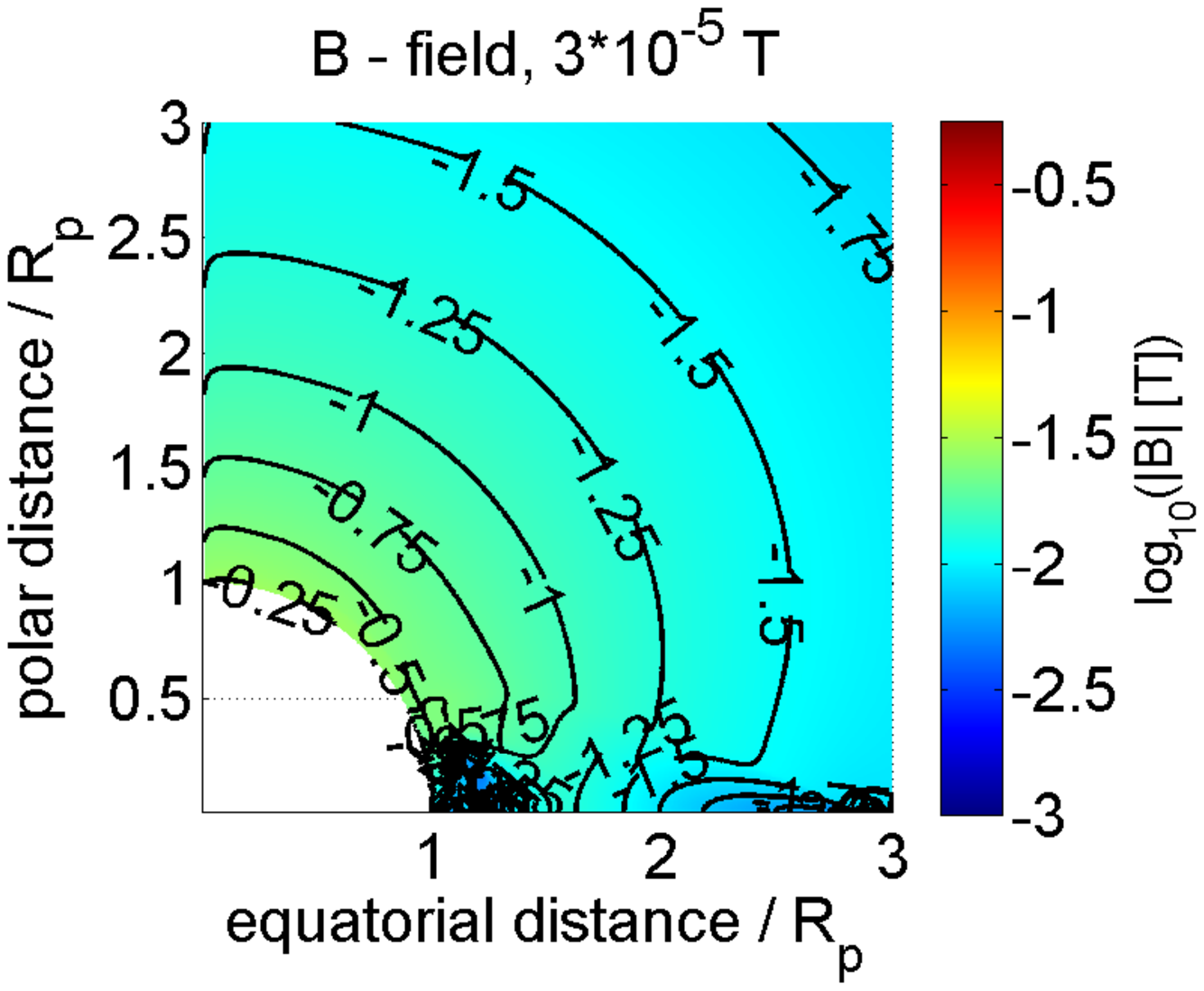}
\includegraphics[width=0.6\columnwidth]{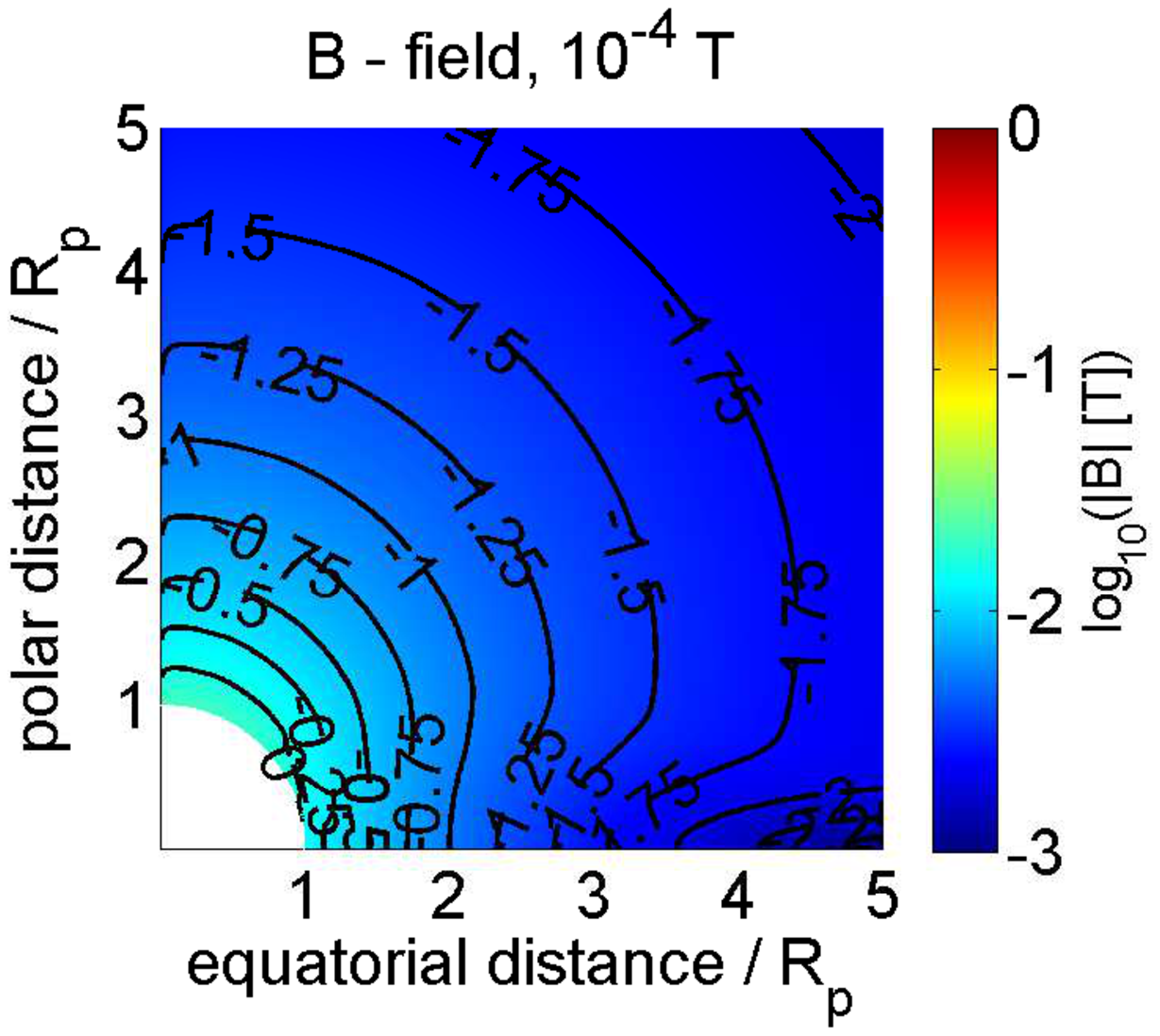}
\caption{Magnetic field of HD 209458b \citep{Khodachenko2015} for 3 different values of equatorial surface magnetic field, $10^{-5}$, $3 \cdot 10^{-5}$ and $10^{-4}$ T (0.1, 0.3 and 1 G, respectively). The $x$-axis and $y$-axis are given in planetary radii $R_{\rm p}$, along the equator and along the pole, respectively. The white quarter of a circle indicates the planet. The black lines indicate lines of constant magnetic field strength.}
\label{fig:fig1}
\end{center}
\end{figure*}

\begin{figure*}
\centering
\includegraphics[width=1.00\columnwidth]{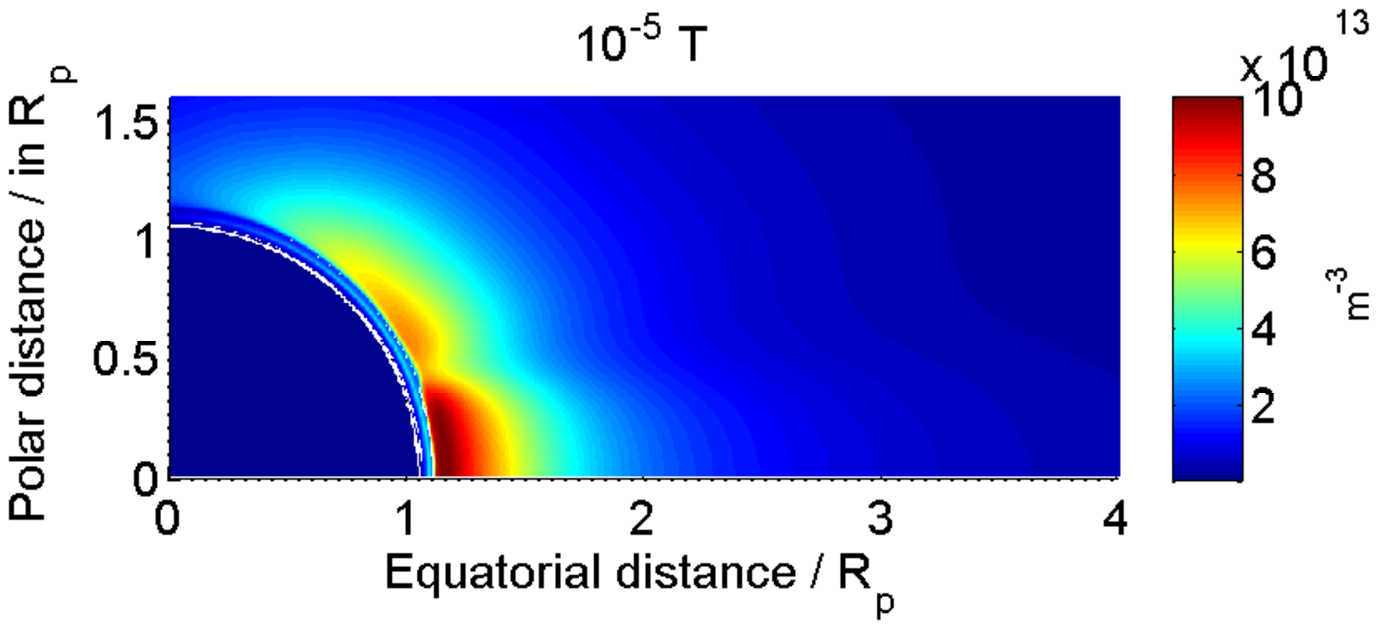}
\includegraphics[width=1.00\columnwidth]{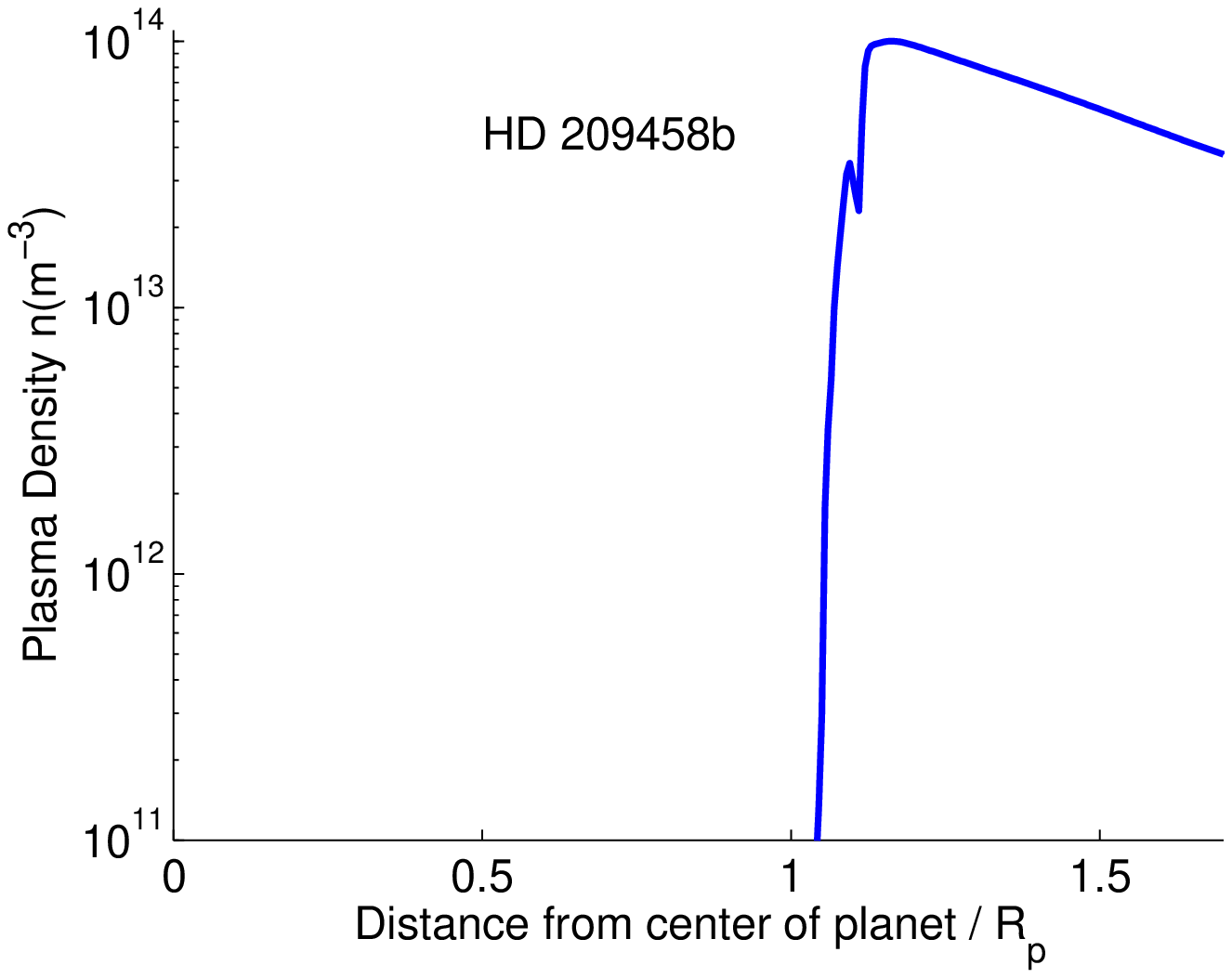}
\includegraphics[width=1.00\columnwidth]{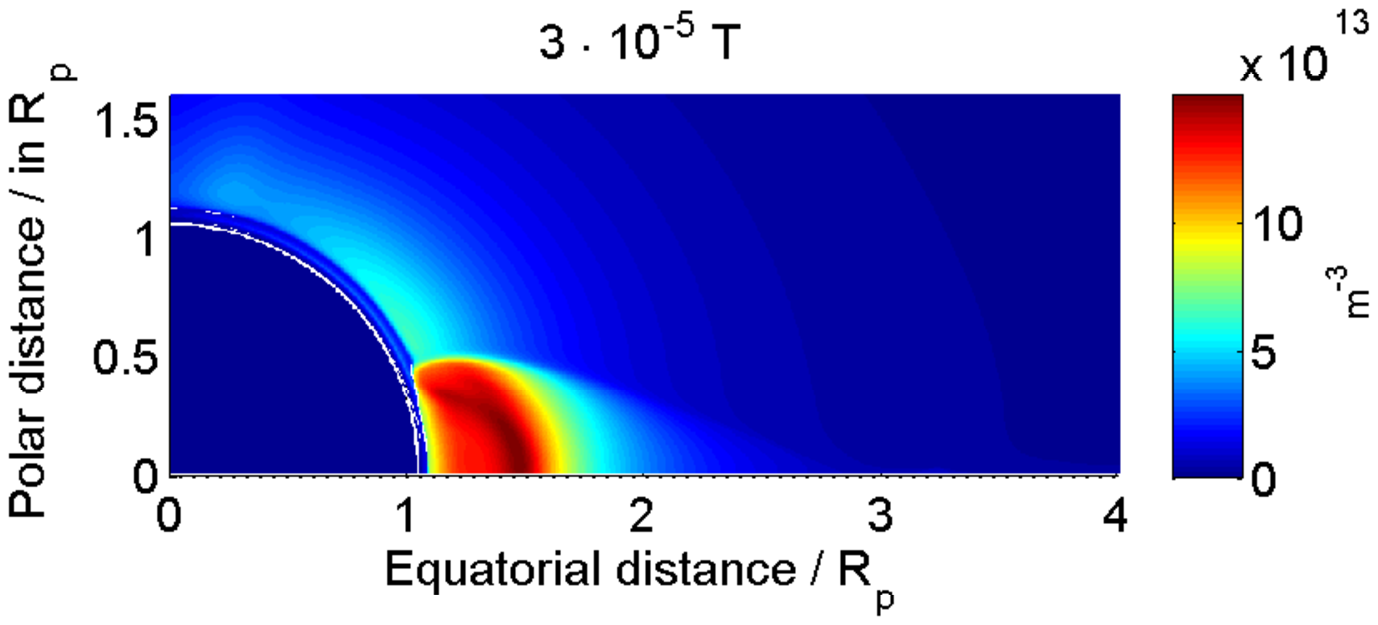}
\includegraphics[width=1.00\columnwidth]{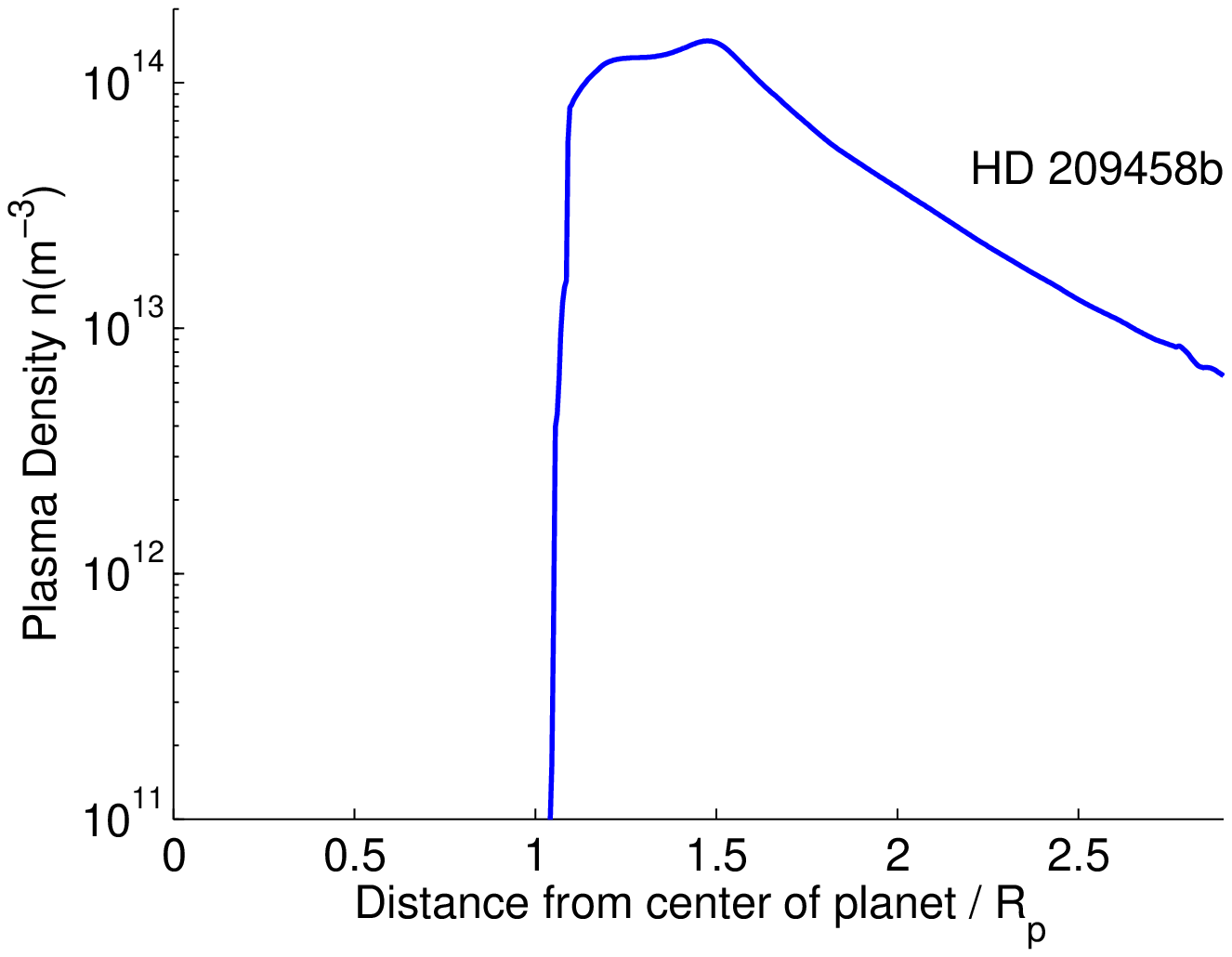}
\includegraphics[width=1.00\columnwidth]{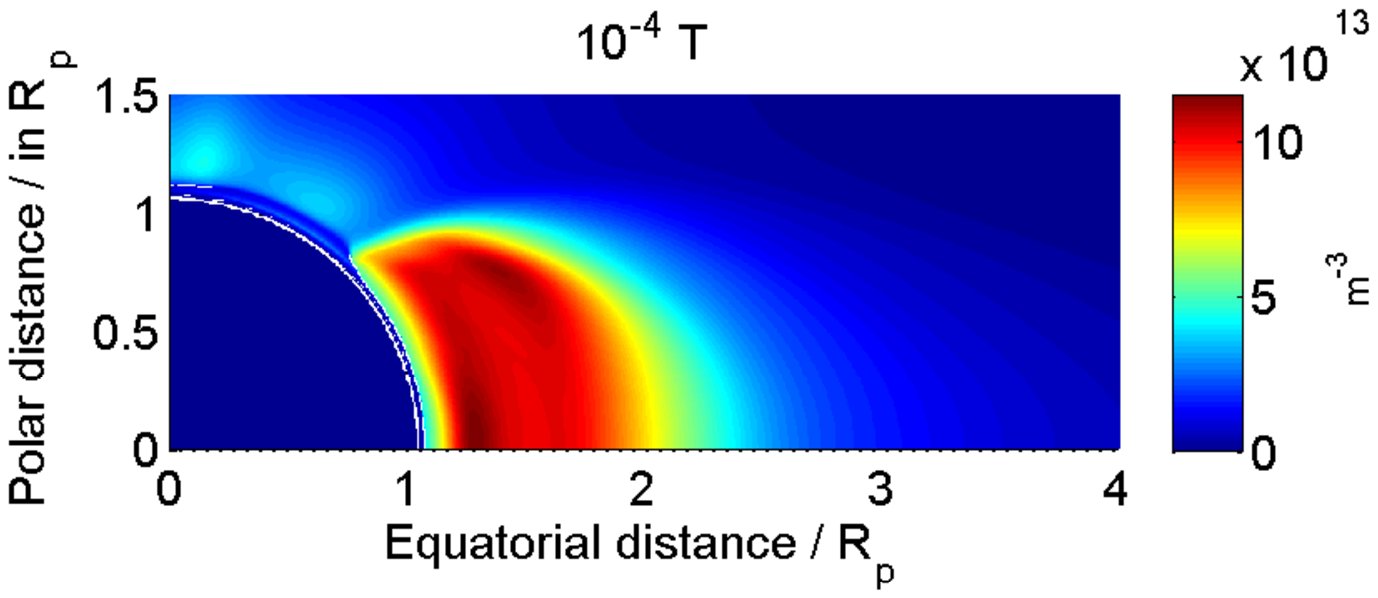}
\includegraphics[width=1.00\columnwidth]{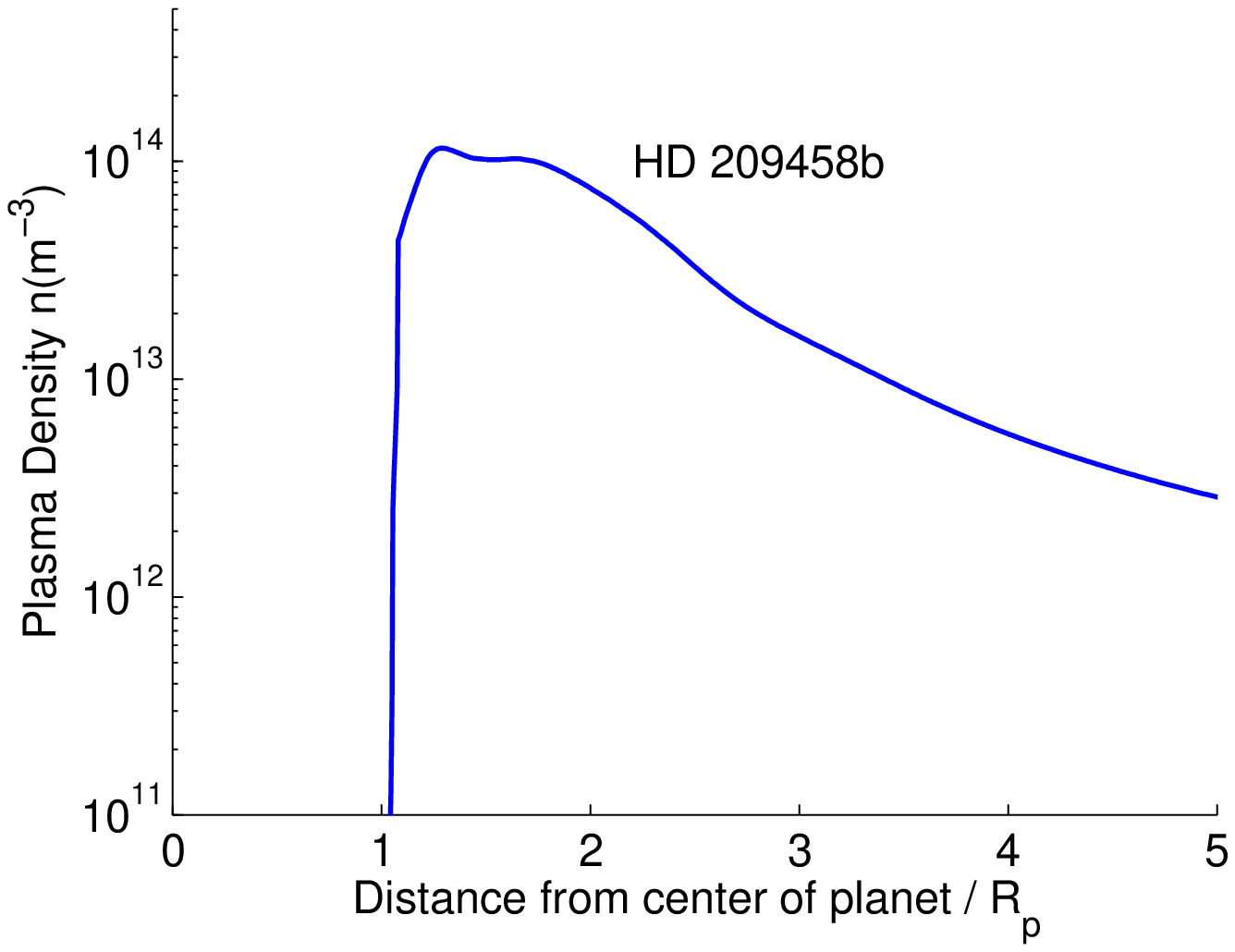}
\caption{Plasma density in $\textnormal{m}^{-3}$ for HD 209458b \citep{Shaikhislamov2014, Khodachenko2015} for the three values of equatorial surface magnetic field. Axes are scaled in planetary radii. The star is located to the right. The plots on the left-hand side are cross sections in the plane spanned by the line connecting the star and the planet (along the equator) and the polar axis. The figures on the right-hand side show corresponding density profiles along the equator towards the star. The profile curves end at the magnetopause. The magnetosphere is filled up with dense plasma up to the magnetopause, a case completely different from and not comparable to Earth or Jupiter.}
\label{fig:fig2}
\end{figure*}

A widely unconstrained but important parameter for obtaining information about radio waves and their escape and generation at exoplanets is the density in the radio emission source region. As mentioned above besides a strong magnetic field the best CMI-condition is also related to a low background plasma within the planetary magnetosphere. Evidently there are no in situ measurements of the plasma densities at exoplanets like for Earth or Jupiter. \citet{Shaikhislamov2014} and \citet{Khodachenko2015} simulated the densities at HD 209458b up to a distance of 20 planetary radii along the equator and 10 planetary radii along the pole. Other authors studied the ionospheres and upper atmospheres of HD 209458b and HD 189733b and obtained similar results \citep{Yelle2004,Munoz2007,Koskinen2010,Koskinen2013a,Koskinen2013b,Koskinen2014,Koskinen2013,Guo2011,Lavvas2014,Salz2015,Chadney2015,Chadney2016,Erkaev2016}.

\begin{figure*}
\begin{center}
\includegraphics[width=1.5\columnwidth]{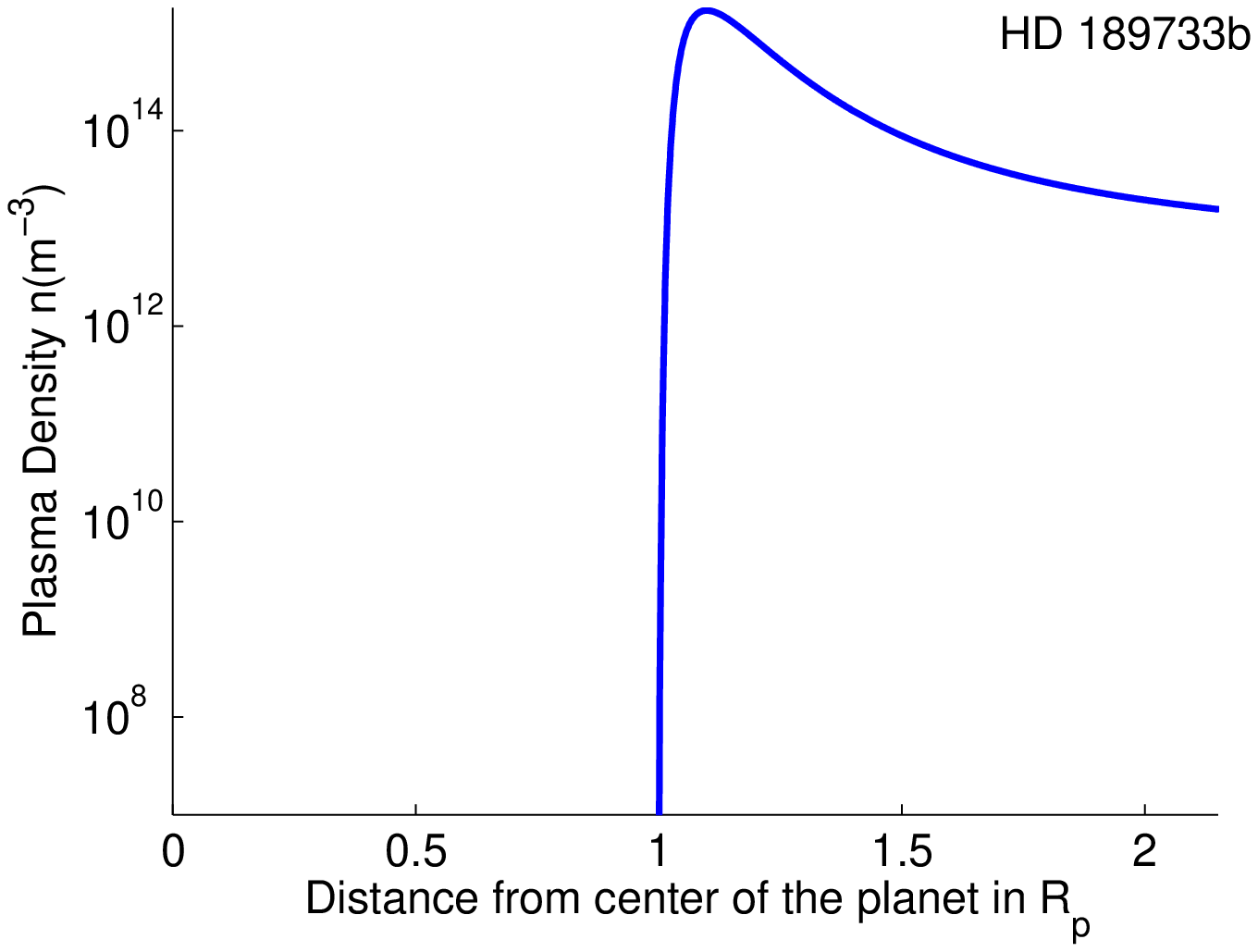}
\caption{Plasma density profile of HD 189733b \citep[adapted from][]{Guo2011}. The $x$-axis is given in planetary radii $R_{\rm p}$, starting from the center of the planet. For HD 189733b the standoff distance is at $\approx 2.2 R_{\rm p}$ for a magnetic moment of $0.3 \mathcal{M}_{\rm J}$. The curve ends at the magnetopause.}
\label{fig:fig3}
\end{center}
\end{figure*}

For the electron density in the atmosphere of HD 209458b we use profiles that have been calculated by the self-consistent axisymmetric 2D magnetohydrodynamics model which is described in detail in \citet{Khodachenko2015}.

Figure \ref{fig:fig1} shows the magnetic field at HD 209458b for three different equatorial surface magnetic field strengths of $10^{-5}$, $3 \cdot 10^{-5}$ and $10^{-4}$ T (0.1, 0.3 and 1 G, respectively). Corresponding magnetic moments are $9.6 \times 10^{25} \textnormal{Am}^2 \approx 0.06 \mathcal{M}_{\rm J}$, $2.9 \times 10^{26} \textnormal{Am}^2 \approx 0.2 \mathcal{M}_{\rm J}$ and $9.6 \times 10^{26} \textnormal{Am}^2 \approx 0.6 \mathcal{M}_{\rm J}$, respectively.

Figure \ref{fig:fig2} shows the corresponding plasma densities at HD 209458b from \citet{Khodachenko2015} for three magnetic field cases, i.e. $10^{-5}$, $3 \cdot 10^{-5}$ and $10^{-4}$ T for the equatorial surface magnetic field of the planet. The left-hand side of the figure shows cross sections in the plane spanned by the line connecting the star and the planet (along the equator) and the polar axis. The right-hand side of Figure \ref{fig:fig2} shows the corresponding electron density profiles of HD 209458b (plotted along the equator towards the star) \citep{Khodachenko2015}. Note that because the magnetic field controls the dynamics of the inner magnetosphere \citep{Khodachenko2015} the plasma densities for the three cases are not the same. The density profile for HD 189733b (Figure \ref{fig:fig3}) is adapted from Figure 3 of \citet{Guo2011} who applied a multi-fluid hydrodynamic upper atmosphere model for the calculation of the planet's neutral and ion profiles as well as its mass loss rate.

As one can see in Figures \ref{fig:fig2} and \ref{fig:fig3} for both Hot Jupiters the magnetosphere is filled up with ionized plasma that hydrodynamically expands up to the magnetopause, i.e. the ionosphere extends out to the magnetospheric boundary. This ionosphere may constitute an obstacle for the propagation of potentially produced radio waves. The standoff distance has been estimated to be at $2.8 R_{\rm p}$ and $2.2 R_{\rm p}$ for  HD 209458b and  HD 189733b, respectively (see Section \ref{sec:sec21}). This upper atmosphere hydrodynamic condition that is caused by the heating of the upper atmosphere due to the host star's powerful XUV radiation at close orbital distances is completely different compared to a hydrostatic upper atmosphere of Jupiter or the Earth. For all solar system planets the upper atmospheres are in the hydrostatic regime where the exobase level is at low altitudes compared to the magnetopause and where there is a wide area of depleted and cool plasma between the exobase and the magnetopause. By applying the electron densities given in Figures \ref{fig:fig2} and \ref{fig:fig3} and the magnetic properties estimated in Section \ref{sec:sec2} we can now investigate under which conditions the CMI can produce radio emissions at extrasolar gas giants. 

\section{Results}
\label{sec:sec4}

Table \ref{tab:tab3} shows a summary of the possibility for the escape and generation of radio emission for different magnetic moment cases studied in this paper in the following sections. The $+$ signs denote that escape or generation is possible and the $-$ signs denote that radio waves cannot escape or be generated. For some cases generation of radio emission might be possible only very close to the planet, indicated by a corresponding note in brackets. The maximum frequency of potentially produced radio emission is given in brackets beside the magnetic moment.

\begin{table*}
\caption{Summary of possibility for generation and/or escape of radio waves for the planets studied in this paper.}
\label{tab:tab3}
\begin{tabular}{|p{4 cm}|p{2 cm}|p{2 cm}|p{2 cm}|p{1.9 cm}|}
 & HD 209458b (pole) & HD 209458b (equator) & HD 189733b (equator) & HD 209458b (1 AU) \\
\hline
\hline
$0.06 \mathcal{M}_{\rm J}$ ($\approx 0.5$ MHz) & generation: $+$ (very close) & generation: $-$ & & \\
 & escape: $-$ & escape: $-$ & & \\
\hline
$0.1 \mathcal{M}_{\rm J}$ ($\approx 0.9$ MHz) & generation: $+$ (very close) & generation: $-$ & & generation: $+$ (very close?) \\
 & escape: $-$ & escape: $-$ & & escape: $-$? \\
\hline
$0.2 \mathcal{M}_{\rm J}$ ($\approx 1.7$ MHz) & generation: $+$ (very close) & generation: $-$ & & \\
 & escape: $-$ & escape: $-$ & & \\
\hline
$0.3 \mathcal{M}_{\rm J}$ ($\approx 4.9$ MHz) & & & generation: $-$ &\\
 & & & escape: $-$ & \\
\hline
$0.6 \mathcal{M}_{\rm J}$ ($\approx 5.6$ MHz) & generation: $+$ (very close) & generation: $-$ & & \\
 & escape: $-$ & escape: $-$ & & \\
\hline
$\mathcal{M}_{\rm J}$ ($\approx 23.9$ MHz) & generation: $+$ (very close) & generation: $+$ (very close) & generation: $+$ (very close) & generation: $+$ (very close?)\\
 & escape: $-$ & escape: $-$ & escape: $-$ & escape: $-$?\\
\hline
$5 \mathcal{M}_{\rm J}$ ($\approx 119.5$ MHz) & generation: $+$ & generation: $+$ & generation: $+$ & generation: $+$\\
 & escape: $-$ & escape: $-$ & escape: $-$ & escape: $+$ \\
\hline
$50 \mathcal{M}_{\rm J}$ ($\approx 1195.1$ MHz) & generation: $+$ & generation: $+$ & generation: $+$ &\\
 & escape: $+$ & escape: $+$ & escape: $-$ & \\
\hline
$100 \mathcal{M}_{\rm J}$ ($\approx 2390.1$ MHz) & generation: $+$ & generation: $+$ & generation: $+$ &\\
 & escape: $+$ & escape: $+$ & escape: $+$ & \\
\hline
\hline
\end{tabular}
\end{table*}

\subsection{Hot Jupiters: HD 209458b and HD 189733b}

\begin{figure*}
\begin{center}
\includegraphics[width=0.7\columnwidth]{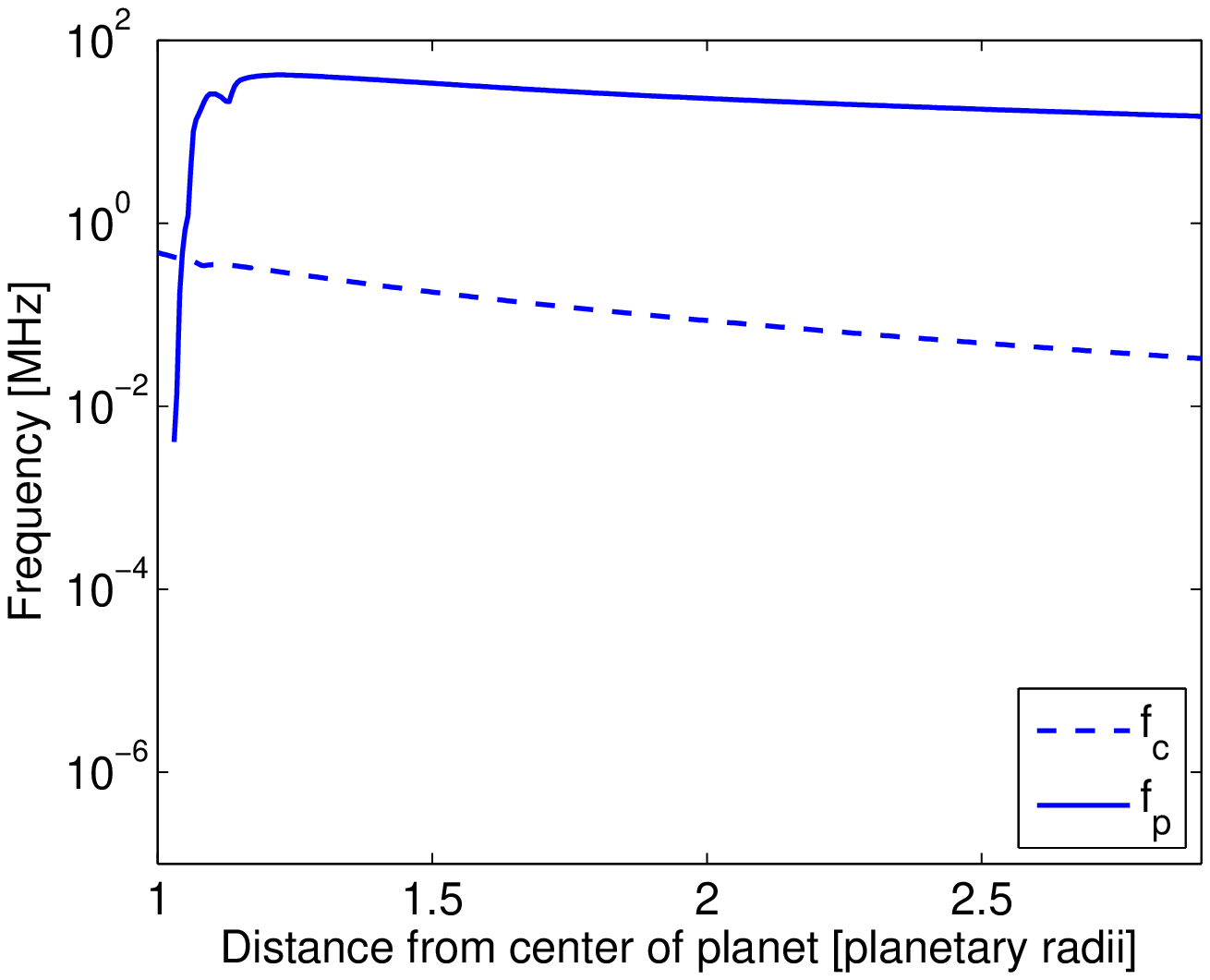}
\includegraphics[width=0.7\columnwidth]{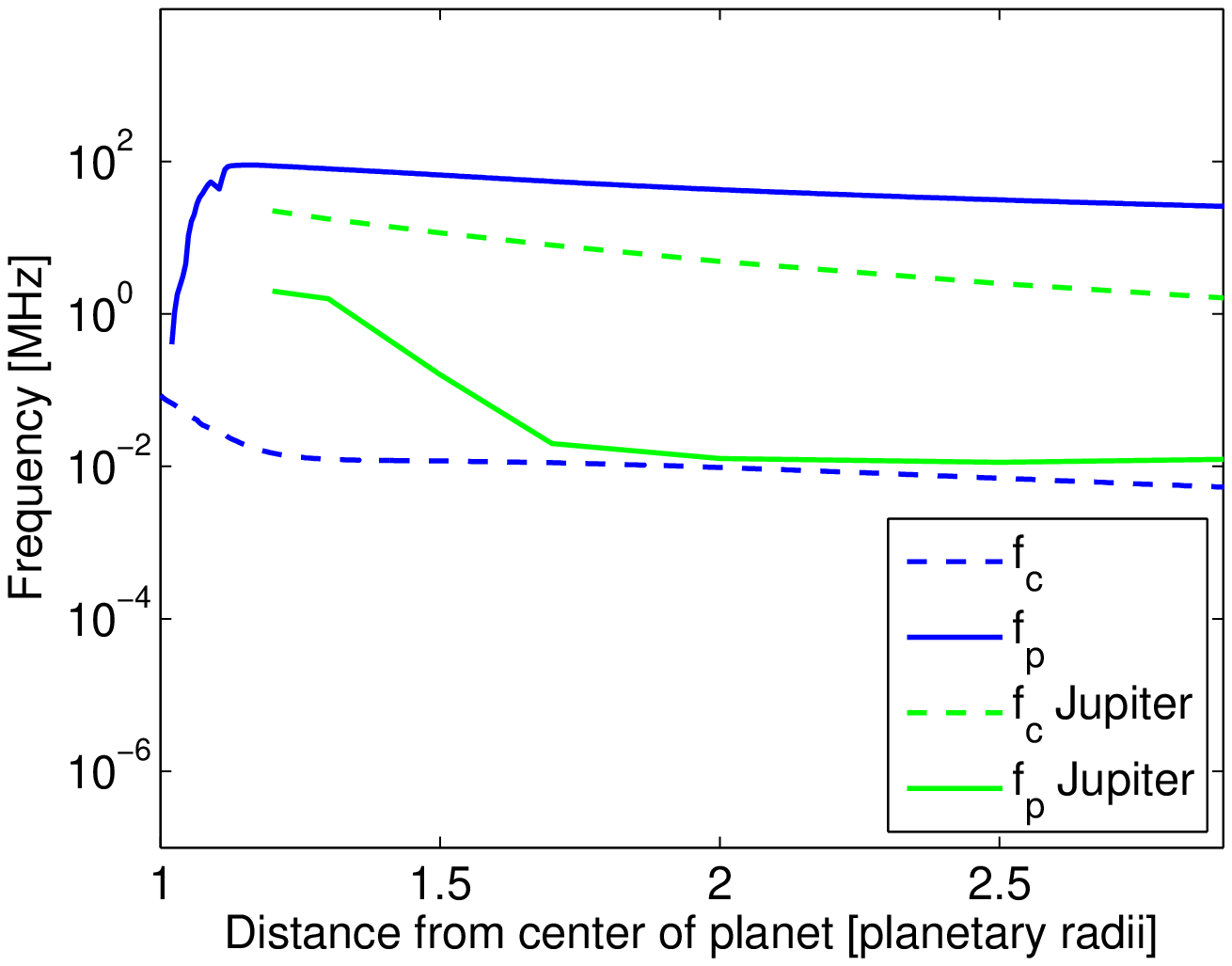}
\includegraphics[width=0.7\columnwidth]{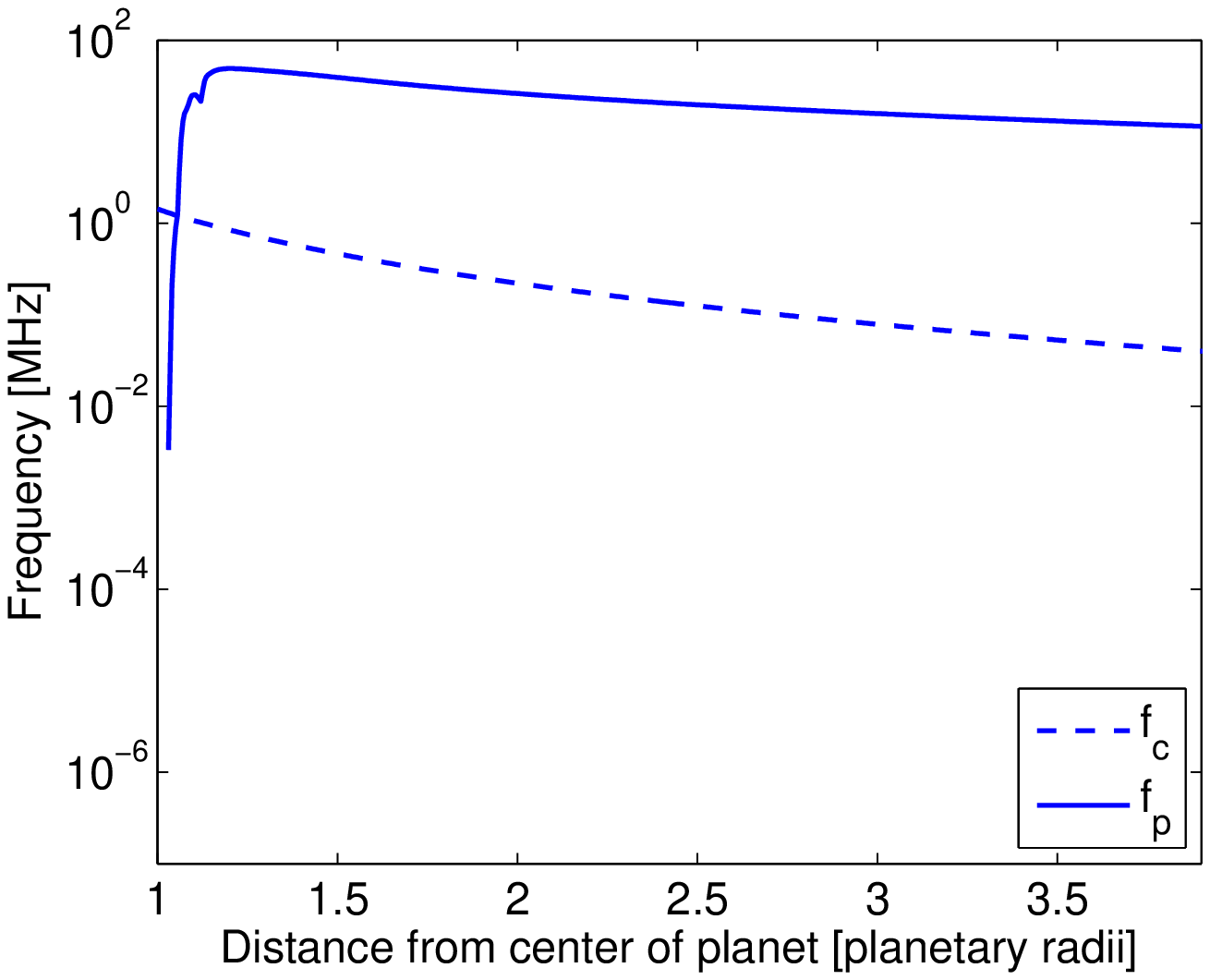}
\includegraphics[width=0.7\columnwidth]{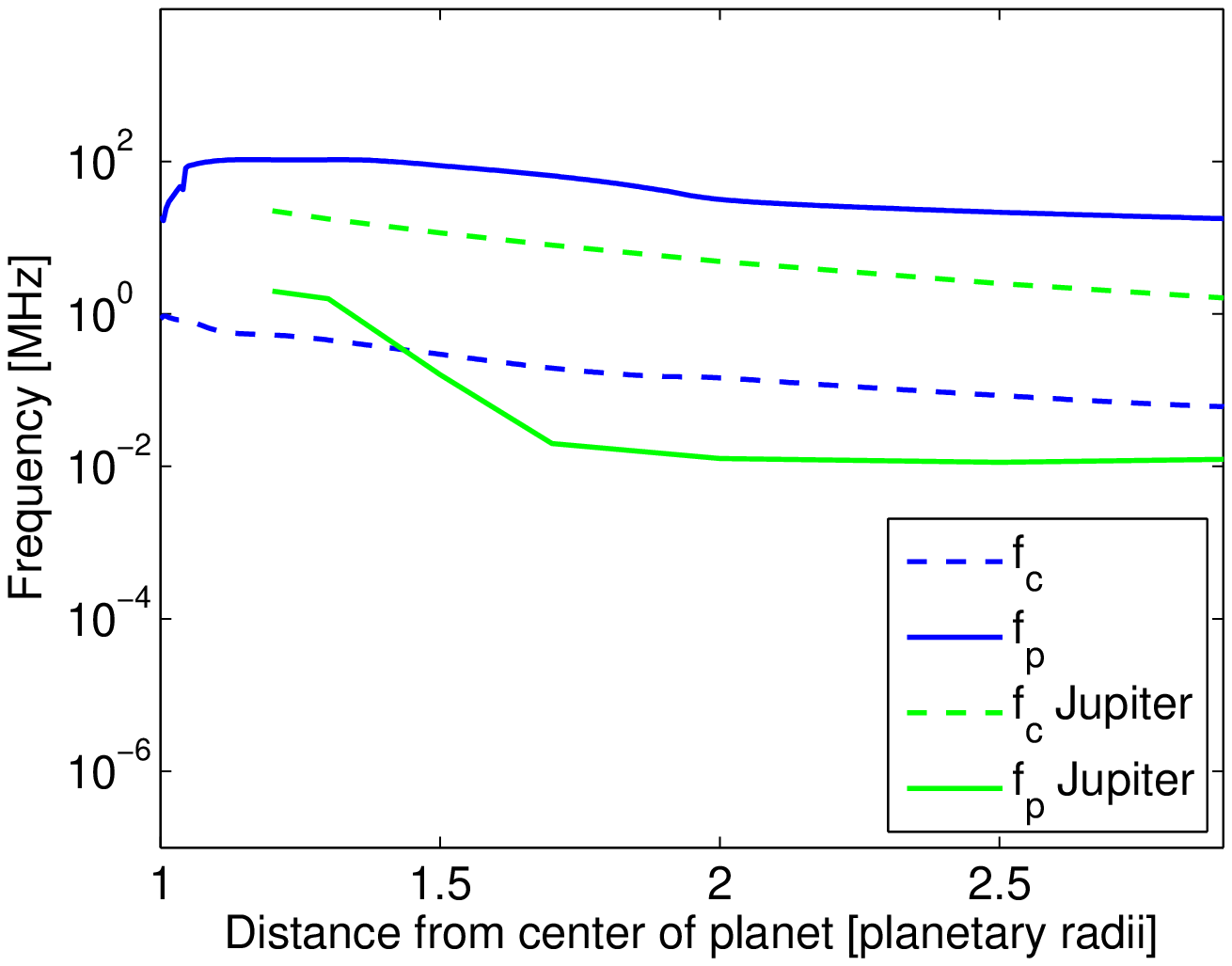}
\includegraphics[width=0.7\columnwidth]{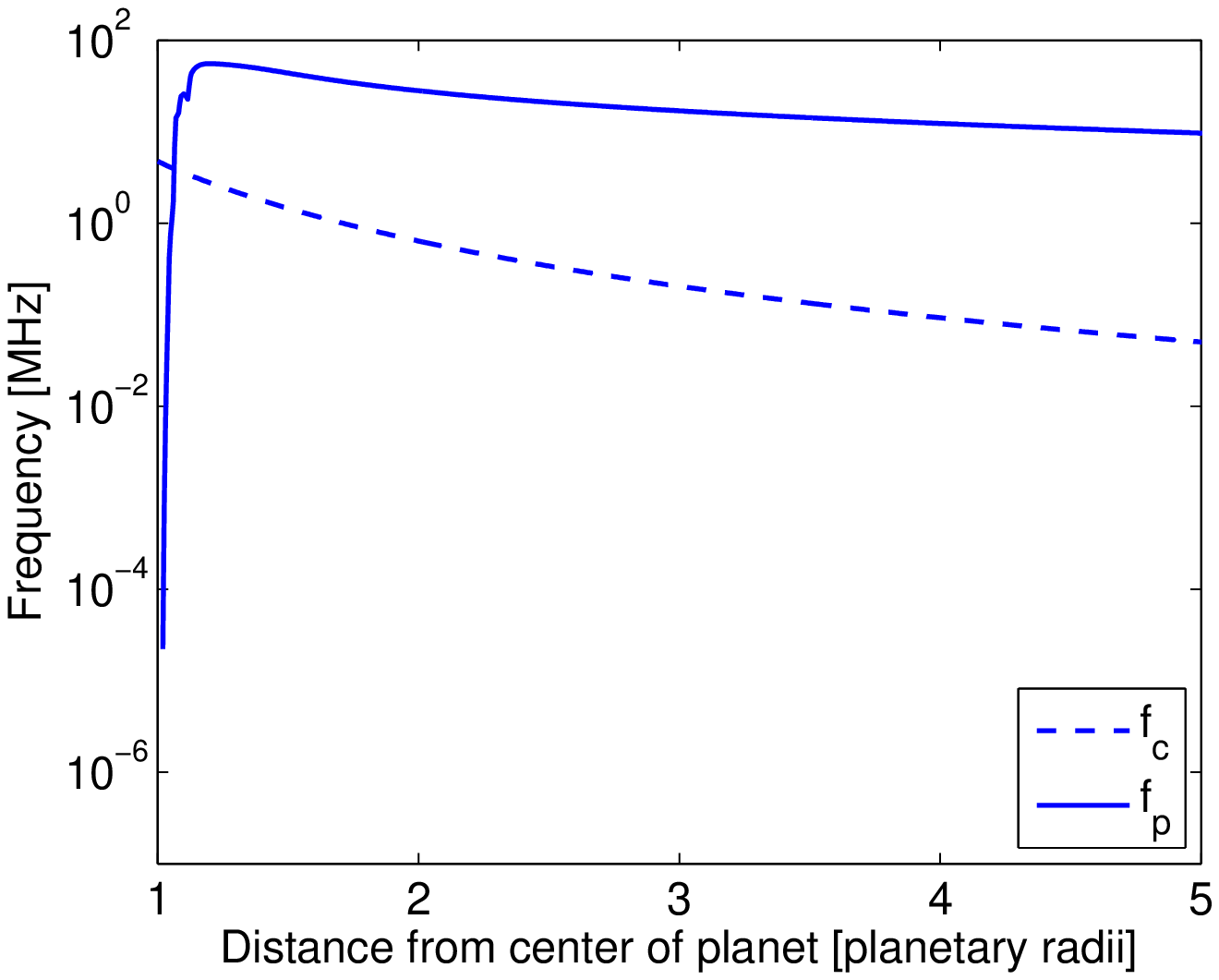}
\includegraphics[width=0.7\columnwidth]{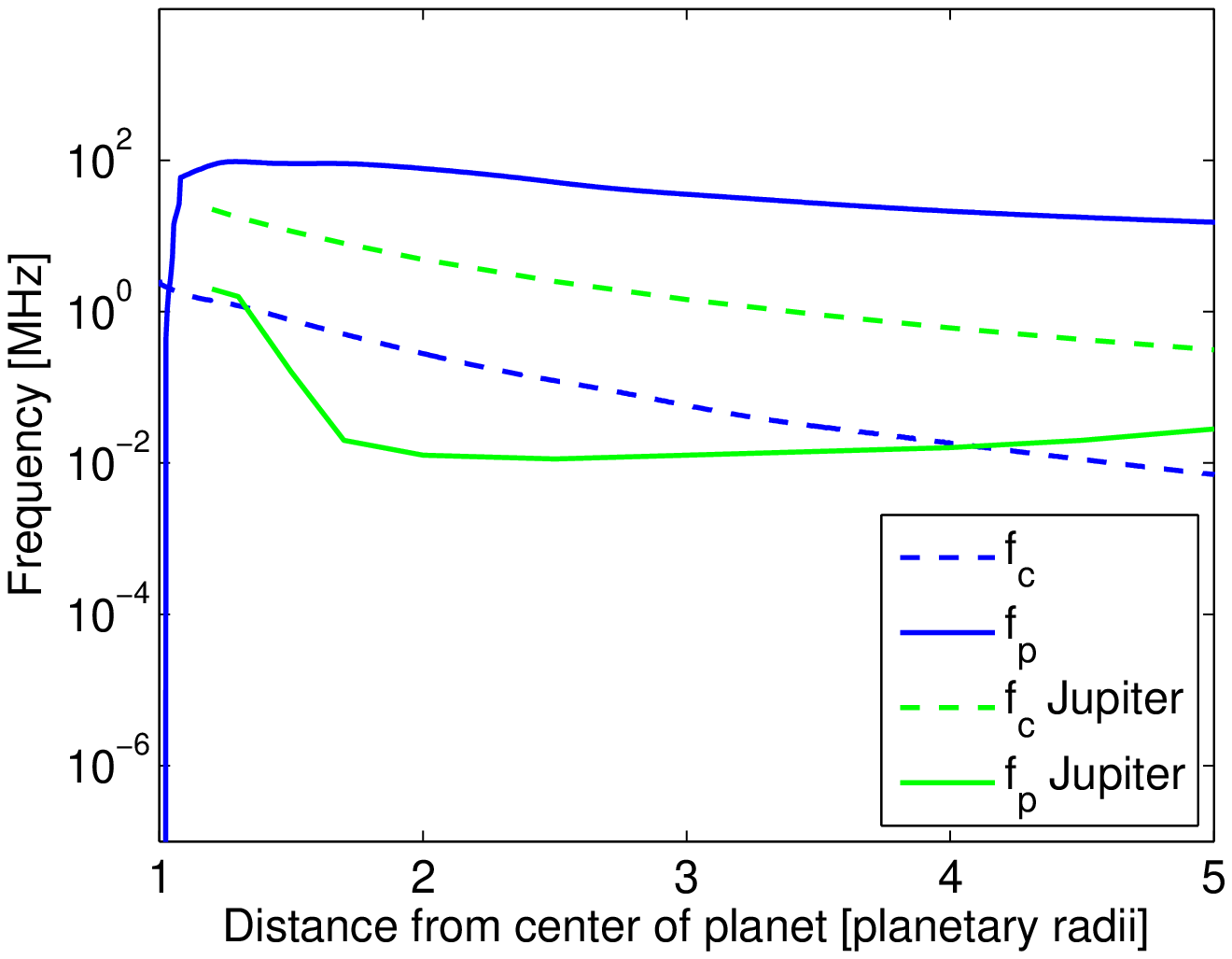}
\caption{Upper panel, left: Plasma frequency $f_{\rm{p}}$ (solid blue line) and cyclotron frequency $f_{\rm{c}}$ (dashed blue line) for HD 209458b for an equatorial surface magnetic field of $10^{-5}$ T as function of distance starting from the planetary transit radius along the pole. Upper panel, right: same as left-hand side but along the equator. The cyclotron frequency and plasma frequency at Jupiter are shown for comparison (green solid and green dashed lines, respectively). The $x$-axes are scaled in units of the respective planetary radii. Middle panel: same as for the upper panel but for an equatorial surface magnetic field of $3 \cdot 10^{-5}$ T. Lower panel: surface magnetic field of $10^{-4}$ T.}
\label{fig:fig4}
\end{center}
\end{figure*}

\begin{figure*}
\begin{center}
\includegraphics[width=0.6\columnwidth]{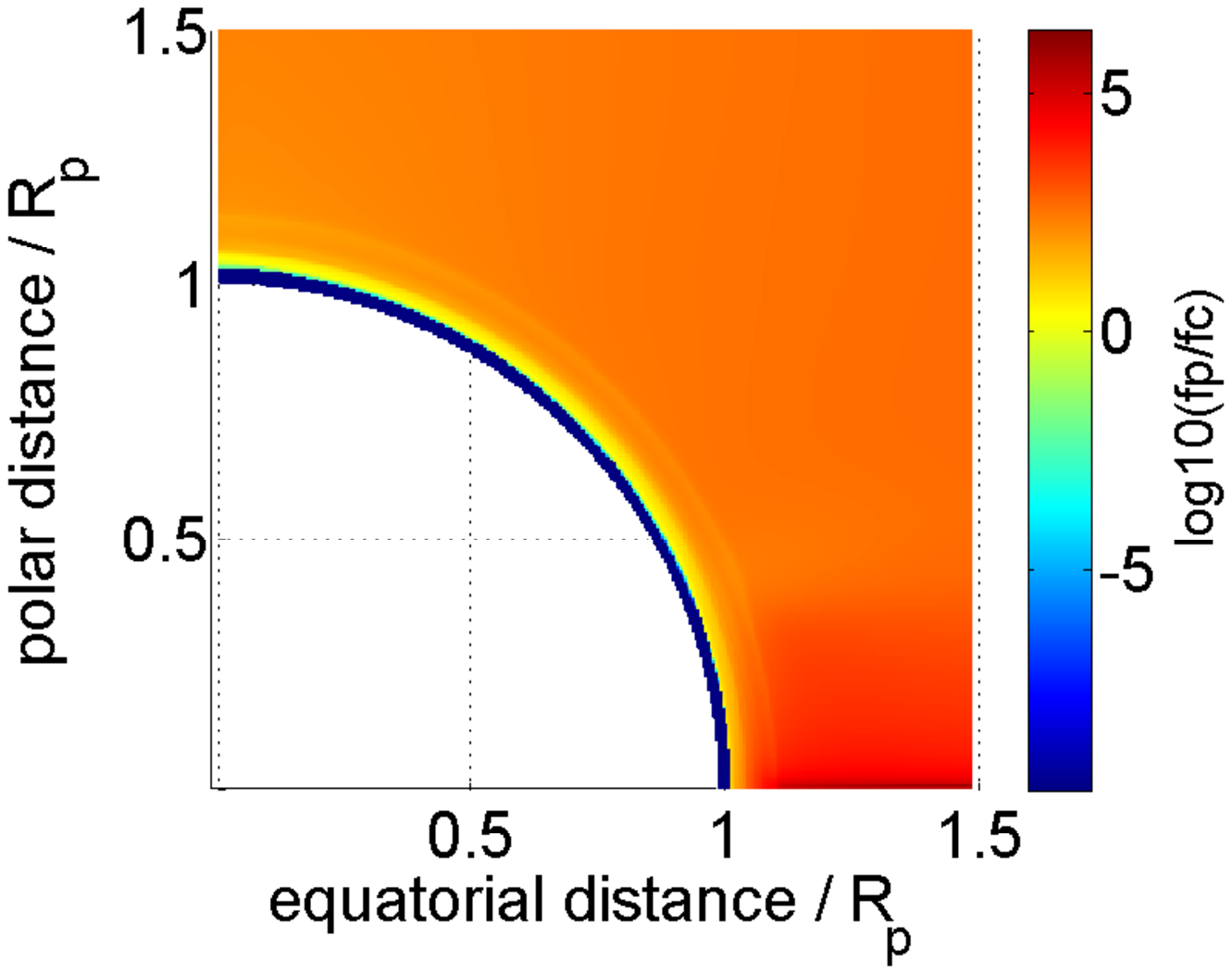}
\includegraphics[width=0.6\columnwidth]{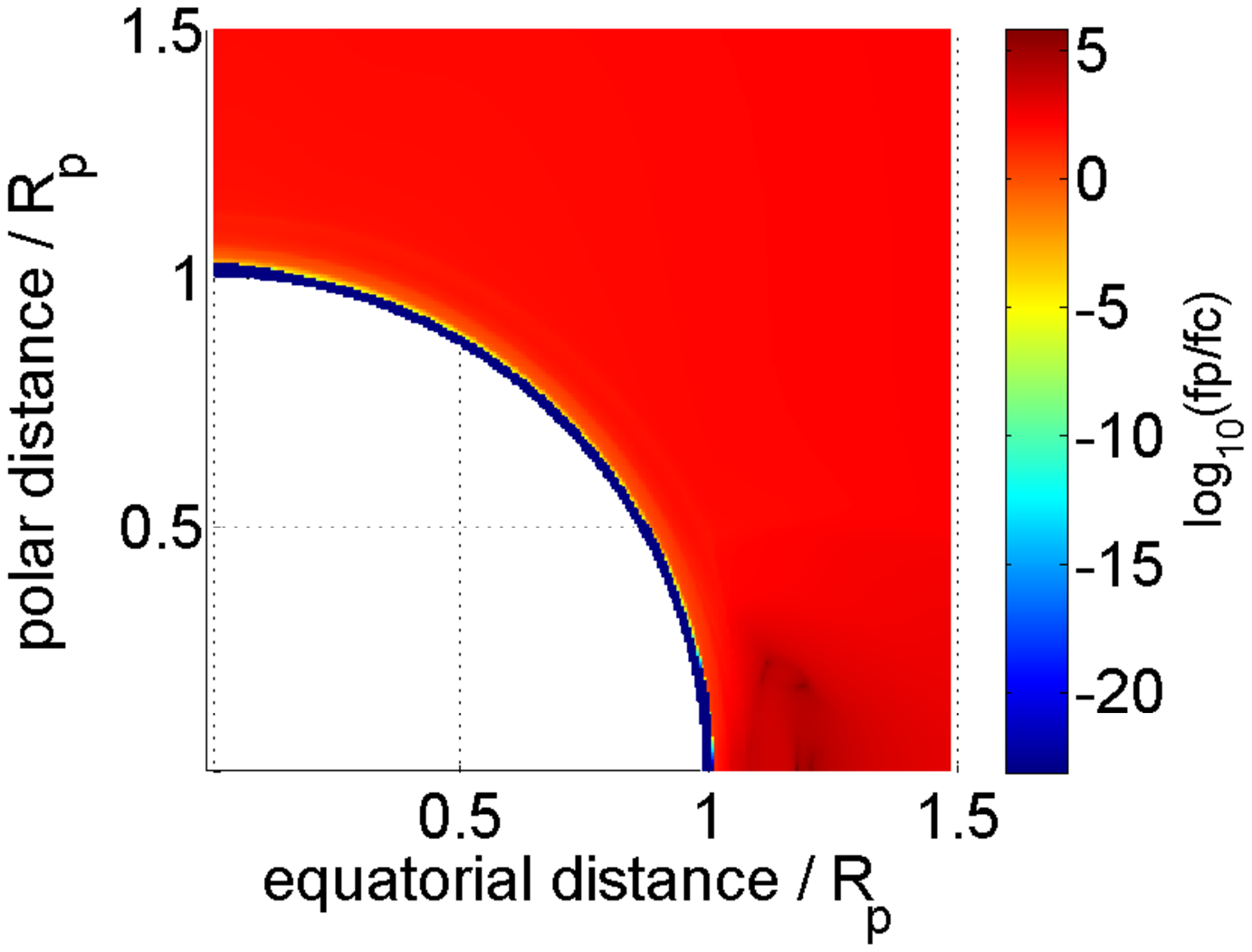}
\includegraphics[width=0.6\columnwidth]{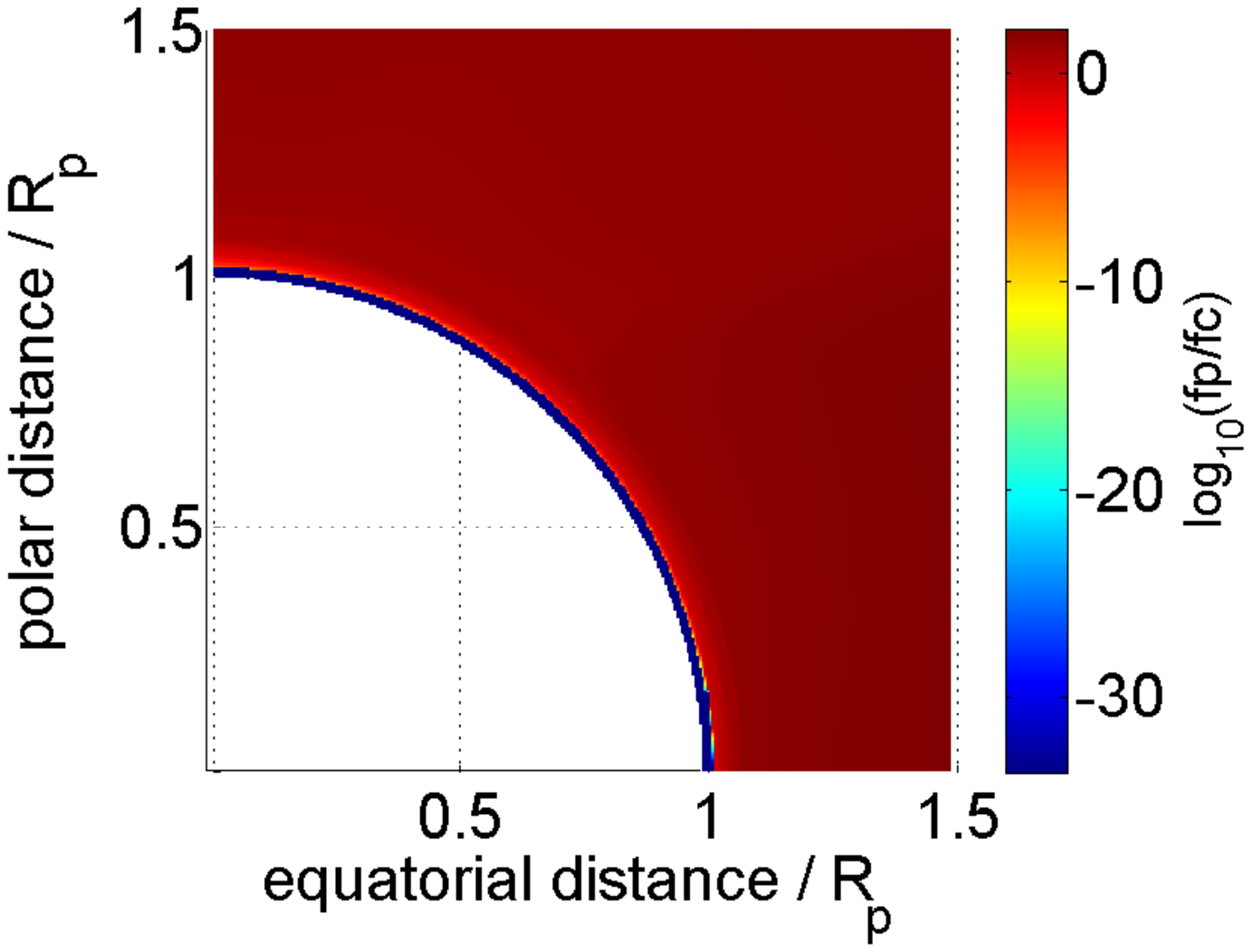}
\caption{Ratios of plasma to cyclotron frequency for equatorial surface magnetic fields of $10^{-5}$, $3 \cdot 10^{-5}$ and $10^{-4}$ T (0.1, 0.3 and 1 G, respectively) for HD 209458b.}
\label{fig:fig5}
\end{center}
\end{figure*}

Figure \ref{fig:fig4} shows the plasma frequency $f_{\rm{p}}$ (solid blue line) and cyclotron frequency $f_{\rm{c}}$ (dashed blue line) for HD 209458b for three different magnetic field cases as function of distance from the planetary surface along the equator towards the star. The top row of panels corresponds to an equatorial surface magnetic field strength of $10^{-5}$ T, the middle row of panels to $3 \cdot 10^{-5}$ T and the lower row of panels to $10^{-4}$ T. The solid green (plasma frequency) and dashed lines (cyclotron frequency) show the Jovian case. The plasma frequency at Jupiter was calculated from plasma density values obtained from \citet{Hess2010b}. Each row of panels shows from left to right the frequencies along the pole and along the equator up to the magnetopause standoff distance. 

For HD 189733b (and also for the cases with higher magnetic moment of the HD 209458b-like planet of Section \ref{sec:josh}) the cyclotron frequency is calculated depending on the distance from the planet assuming the magnetic field to be dipolar, i.e. neglecting fields generated by magnetopause currents and higher field harmonics. We used Equation (\ref{eqn:freqbrelation}) to calculate the cyclotron frequency. 

For HD 209458b with a magnetic moment of $0.06 \mathcal{M}_{\rm J} - 0.6 \mathcal{M}_{\rm J}$ generation of radio emission might be possible at the pole and very close to the planet, i.e. below $0.05 R_{\rm p}$, but escape of the radio waves would not be possible (see Figure \ref{fig:fig4} and Table \ref{tab:tab3}). A question left open for follow-up studies is whether the energetic electrons needed for the generation of CMI-induced radio emission are present at such close distances to the planet's surface. The frequencies at the equator - where the magnetic field is only half as large as at the pole - are such that the generation and escape of radio waves is inhibited (see left-hand side of Fig. \ref{fig:fig4} and Table \ref{tab:tab3}).
 
\begin{figure*}
\begin{center}
\includegraphics[width=1.00\columnwidth]{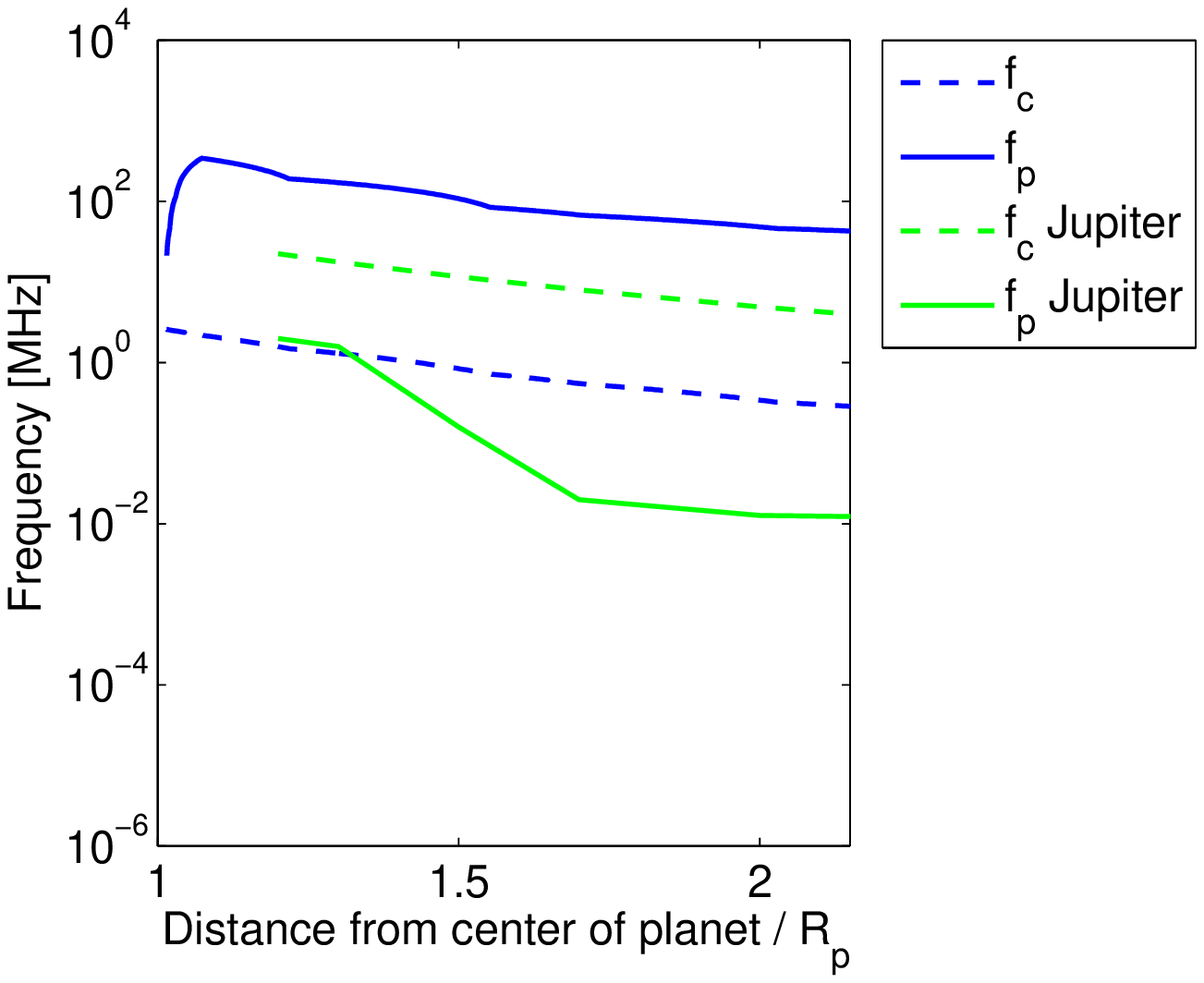}
\caption{Left: Plasma frequency $f_{\rm{p}}$ (solid blue line) and cyclotron frequency $f_{\rm{c}}$ (dashed blue line) for HD 189733b as function of distance from the planetary surface along the equator, starting from the planetary transit radius. $f_{\rm{c}}$ corresponds to $0.3 \mathcal{M}_{\rm J}$. The cyclotron frequency and plasma frequency at Jupiter are shown for comparison (green solid and green dashed line, respectively).}
\label{fig:fig6}
\end{center}
\end{figure*}

Figure \ref{fig:fig6} shows the plasma (solid blue line) and cyclotron frequency (dashed blue line) for HD 189733b with $0.3 \mathcal{M}_{\rm J}$, i.e. a higher magnetic moment than for HD 209458b. The cyclotron frequency is below the plasma frequency from the planet up to the magnetopause. 

Figure \ref{fig:fig5} shows the ratio of plasma to cyclotron frequencies for the three equatorial magnetic field strengths of $10^{-5}$, $3 \cdot 10^{-5}$ and $10^{-4}$ T, from left to right. One can see that the region of possible generation of radio waves via the CMI lies very close to the planet. Here we only study weak magnetic fields in detail but preliminary results from a follow-up study on Tau Bootis b, a very massive Hot Jupiter with a mass of $M \sin i = 4.13 M_{\rm J}$ (from \url{http://exoplanet.eu/}) and a magnetic moment of $0.76 \mathcal{M}_{\rm J}$ \citep{Griessmeier2005,Griessmeier2007c,Griessmeier2007b}, already indicate that this planet might have the same problems at its orbital distance of about 0.045 AU but probably not at slightly larger orbits \citep{Weber2017}.

Since for the cases discussed in this section the planetary surface magnetic field is such that corresponding frequencies of radio emission would be below the Earth's ionospheric cutoff, we discuss higher magnetic field cases in the next section (\ref{sec:hm}).

\subsection{Influence of higher magnetic field}\label{sec:hm}

\subsubsection{HD 209458b}\label{sec:hm1}

We also checked a young and massive planet with corresponding high magnetic moments of $50 \mathcal{M}_{\rm J}$ and $100 \mathcal{M}_{\rm J}$, respectively, for the same plasma conditions as for HD 209458b (see Table \ref{tab:tab3}). Such high magnetic moments could possibly be generated by very young, rapidly rotating, massive planets. We adopt these values as maximum expected planetary magnetic moments. $50 \mathcal{M}_{\rm J}$ would correspond to an exoplanet with an age of 100 Myr and 13 Jupiter masses and $100 \mathcal{M}_{\rm J}$ to a planet with age of 10 Myr and 13 Jupiter masses in the model of \citet{Reiners2010}. These cases are not shown in the figures because these estimations have to be taken with care. The magnetic field clearly controls the plasma dynamics in the inner magnetosphere \citep{Khodachenko2015}, thus the whole problem should be treated self-consistently. For the case of HD 209458b (Figures \ref{fig:fig1}, \ref{fig:fig2}, \ref{fig:fig4} and \ref{fig:fig5}) we performed these self-consistent calculations. Due to lack of data this was not possible for HD 189733b, the HD 209458b-like planet of Section \ref{sec:josh} and the high magnetic moment cases of the current section. However, changes in the order of a factor ten of plasma densities or magnetic field do not alter the overall result. If the plasma conditions were the same as for HD 209458b for the high magnetic field of $50 \mathcal{M}_{\rm J}$ and $100 \mathcal{M}_{\rm J}$ generation as well as escape of radio waves would be possible at the pole and at the equator whereas for $\mathcal{M}_{\rm J}$ and $5 \mathcal{M}_{\rm J}$ escape would be inhibited (see Table \ref{tab:tab3}).

\subsubsection{HD 189733b}\label{sec:hm2}

If the plasma conditions of HD 189733b are considered then for $\mathcal{M}_{\rm J}$, $5 \mathcal{M}_{\rm J}$ and $50 \mathcal{M}_{\rm J}$ generation of radio waves would be possible but they cannot overcome the barrier of plasma frequency. Only for $100 \mathcal{M}_{\rm J}$ generation and escape of radio emission would be possible (see Table \ref{tab:tab3}). Note that for HD 189733b we only consider locations in the equatorial plane. 

Massive objects with such high magnetic moments might have a much more compact atmosphere and, therefore, could have a region filled with diluted plasma. This might enable escape of radio emission also for $50 \mathcal{M}_{\rm J}$ but further analysis in follow-up studies is required. 

The model from \citet{Reiners2010} predicts a polar magnetic field strength of 0.0014 T (14 G, $\approx 2.4 \mathcal{M}_{\rm J}$) for HD 189733b (with planetary mass of $1.13 M_{\rm J}$ and an age of 1.7 Gyr), i.e. a similar value as for the maximum polar magnetic field of Jupiter. So the dashed green line of the cyclotron frequency at Jupiter in Figure \ref{fig:fig6} corresponds roughly to the cyclotron frequency of HD 189733b when tidal locking has no influence on the magnetic field. It is about a factor of ten below the plasma frequency of HD 189733b as modelled by \citet{Guo2011}. Even though they considered the non-magnetized case in their model, with a factor of ten difference between plasma and cyclotron frequency the overall results, i.e. the fact that radio waves cannot be produced and cannot escape from the source region, should remain unchanged.

\subsubsection{General case}\label{sec:hm3}

We have shown that even for much larger magnetic moments than expected for the considered planets the ratio of plasma versus cyclotron frequency lies above 0.4 at distances larger than $\sim 0.05 R_{\rm p}$. For both planets -- HD 209458b and HD 189733b -- the cyclotron frequency is higher than the plasma frequency only very close to the planets. We found that, starting from $10^4 \mathcal{M}_{\rm J}$, the plasma frequency is below the cyclotron frequency up to 4 planetary radii and beyond. This magnetic moment is unrealistically high. Nevertheless, at least for brown dwarfs the CMI has been confirmed to be the dominant generation process for radio emission \citep{Hallinan2008}. \citet{Hallinan2008} found emission which requires magnetic fields in the kG range. Due to their high mass brown dwarfs should also have much more compact atmospheres which might also contribute to the fact that they have the CMI mechanism operating. Our results indicate that with an unrealistic (for a Jupiter-like planet) kG-field the conditions for the CMI are fulfilled for HD 189733b and close to the planet for HD 209458b. But even for a large planetary magnetic field the question remains if the produced radio waves can escape from the planet through the dense plasma. However, the magnetic field influences the mass loss from the planet, thus a higher magnetic field could inhibit the filling of the magnetosphere with dense plasma. But close-in planets with such high magnetic fields (high enough to alter our overall results) are not known. Amongst the Hot Jupiters, Tau Bootis b is already one of the planets with the strongest expected magnetic field at such close orbital distances \citep{Griessmeier2005,Griessmeier2007c,Griessmeier2007b}.

It is very likely that such strong magnetic moments ($50 \mathcal{M}_{\rm J}$ or $100 \mathcal{M}_{\rm J}$) for Hot Jupiters are not realistic. Even if one assumes that the magnetic moment is larger than the one estimated by the reproduction of the Ly-$\alpha$ HST observations during HD 209458b's transit, the magnetic moments of Hot Jupiters are expected to be too weak and the ionization degrees and plasma densities in their upper atmospheres up to the magnetopause boundaries at these close orbital locations are too high so that one can expect that the CMI cannot operate under such extreme conditions. As one can see from Figures \ref{fig:fig2} and \ref{fig:fig3} for both planets the magnetosphere is filled up with dense plasma, i.e. the ionosphere extends out to the magnetospheric boundary which constitutes an obstacle for the propagation of potentially produced radio waves. Realistic changes of the density conditions would not lead to more favourable results for planets like HD 209458b or HD 189733b. A plasma density which is $10^{-8}$ lower would be required to fulfill the condition $f_{\rm{p}}/f_{\rm{c}} < 0.4$ for a Hot Jupiter like HD 209458b.

Including magnetopause currents, the magnetic field at the magnetopause is higher than the pure dipolar model by a small factor $\eta$. For example, for the magnetospheric model of \citet{Griessmeier2004,Griessmeier2005,Griessmeier2007b}, $\eta = 2.32$. As can be seen in Figure \ref{fig:fig4}, a change of the magnetic field by a factor $< 10$ does not change the fact that $f_{\rm{p}}/f_{\rm{c}} >> 0.4$.

We note again that for HD 189733b and for the other higher magnetic field cases we have not performed the self-consistent calculation including an intrinsic planetary magnetic field as for HD 209458b. Such simulations exist up to now only for HD 209458b \citep{Khodachenko2015} and they are very time consuming. But as shown above the effect of the intrinsic planetary magnetic field on plasma densities in the magnetospheric environment is not so strong that it alters the qualitative results.

\subsubsection{Comparison to the solar system}\label{sec:hm4}

For Jupiter the cyclotron frequency is higher than the plasma frequency in the whole plotting range (see e.g. Figure \ref{fig:fig4}) which would not be the case for $50 \mathcal{M}_{\rm J}$ and $100 \mathcal{M}_{\rm J}$, where the plasma frequency is lower than the cyclotron frequency only below $\approx 2$ planetary radii. This shows how dense the plasma in the magnetosphere of HD 209458b at 0.047 AU is compared to Jupiter at 5.2 AU.

For Jupiter at 5.2 AU, the situation is completely different and there are large regions with dilute plasma, i.e. with a low plasma frequency. Because Jupiter's upper atmosphere and ionosphere are hydrostatic, the exobase level is far below the magnetopause distance \citep{Yelle1996}. Up to about 5 planetary radii the ratio of plasma to cyclotron frequency is below $0.3$. The solid and dashed green lines in Figures \ref{fig:fig4} and \ref{fig:fig6} show the Jovian case. The density or plasma frequency and thus ratios of plasma to cyclotron frequency start to rise after a minimum around 1.7 planetary radii which comes from the fact that the Io plasma torus with enhanced plasma density is located at about 5.9 Jovian radii. It is possible that such a moon is also orbiting Eps Eridani b or Jovian-like exoplanets in general, a hypothesis which has been investigated by \citet{Nichols2011} and \citet{Noyola2014}. Also for Earth, the gyrofrequency lies above the plasma frequency from about 1 planetary radius up to 6 Earth radii \citep[see Figure 13 in][]{Gurnett1974}. However, for Hot Jupiters an Io-like plasma source is probably impossible, because they likely can have none or at least only tiny moons (e.g., \citet{Kislyakova2016} and references therein).  On the other hand, \citet{Kislyakova2016} have shown that for the hottest planets, a Trojan swarm orbiting in 1:1 resonance with the giant planet could provide a plasma source, but further investigation of CMI conditions in these systems is needed.

The upper atmospheres of Hot Jupiters experience hydrodynamic outflow up to the magnetopause, which is related to the high XUV radiation and ionization degrees. This produces a highly unfavourable environment so that the CMI most likely cannot operate and radio waves cannot be emitted. The main difference between Hot Jupiters and radio emitting planets in the solar system is that solar system planets have hydrostatic upper atmospheres that result in much better conditions for the generation of the CMI and the emission of radio waves. Therefore, we investigate in the next section at which orbital location around a solar-like host star the upper atmosphere of an extrasolar gas giant may undergo the transformation from the hydrodynamic to the hydrostatic regime.

\subsubsection{Outlook}\label{sec:hm5}

The influence of higher magnetic fields on the results needs closer investigations which will be performed in follow-up studies. One of these studies will be done for Tau Bootis b, because it is much more massive than HD 209458b and HD 189733b and it's magnetic moment is predicted to be higher with corresponding radio emission above the ionospheric cutoff \citep{Griessmeier2005,Griessmeier2007c,Griessmeier2007b}. The preliminary results of this study are much more promising than for HD 209458b and HD 189733b \citep{Weber2017}. 

\subsection{Transition region from hydrodynamic to hydrostatic - HD 209458b-like planets between 0.2--1 AU}\label{sec:josh}

\citet{Chadney2015,Chadney2016} recently studied the upper atmosphere structure, escape rate and the electron density for an HD 209458b-like gas giant orbiting between 0.2 and 1 AU from its host star. The studies were performed with a hydrodynamic upper atmosphere model, which includes hydrogen photochemistry, heating and cooling processes as well as stellar XUV radiation input. The main aim of their study was to investigate under which stellar XUV flux conditions the atmospheric mass loss of Hot Jupiters will change from the hydrodynamical to the Jeans escape regime. It has to be mentioned that this atmosphere model does not include the effect of the planetary magnetic field, i.e. it considers a non-magnetized case, in contrast to the model by \citet{Khodachenko2015}. 

According to \citet{Chadney2015}, the transition region from hydrodynamic to hydrostatic conditions occurs between 0.2 and 0.5 AU for an HD 209458b-like planet around a Sun-like star. Close to the star the planetary upper atmosphere expands hydrodynamically and at orbital distances greater than 0.5 AU the atmosphere is in hydrostatic equilibrium. In the latter case, the exobase level is very close to the planet while the magnetopause standoff distance moves also to larger distances because of the decrease in ram pressure of the stellar wind and the increase of the planetary magnetic moment. In the hydrostatic regime the exobases are located close to the planetary radius $R_{\rm p}$ (similar to solar system planets). For an HD 209458b-like planet at 1 AU around a Sun-like star the exobase altitude is at about $\approx 1.05 R_{\rm p}$. At 0.5 AU the exobase altitude is $\approx 1.72 R_{\rm p}$, i.e. still close to the planet, while at 0.2 AU the planet is already in the hydrodynamic regime and the exobase is located above the model domain (which extends to $16 R_{\rm p}$), i.e. one can expect a magnetosphere filled up with dense plasma up to the magnetopause and no favourable conditions for the CMI. For a star more active than the Sun, like e.g. Eps Eridani, the transition from hydrodynamic to Jeans escape occurs at slightly larger distances of 0.5 - 1 AU \citep[from][]{Chadney2015}.

So for close-in planets the magnetospheric cavity is filled up with dense plasma, but at orbit locations and corresponding XUV fluxes of about 0.5 AU (for a Sun-like star) the conditions for the CMI are already quite favourable for gas giants with about 5 times the Jovian magnetic moment. According to \citet{Griessmeier2007b} planets with such a magnetic moment should exist, e.g. 70 Vir b with $4.5 \mathcal{M}_{\rm J}$ or HD 114762b with $5.1 \mathcal{M}_{\rm J}$. At 1 AU the conditions are comparable to Earth or Jupiter, i.e. above the exobase the density is lower and fast decreasing.

\begin{figure*}
\begin{center}
\includegraphics[width=1.5\columnwidth]{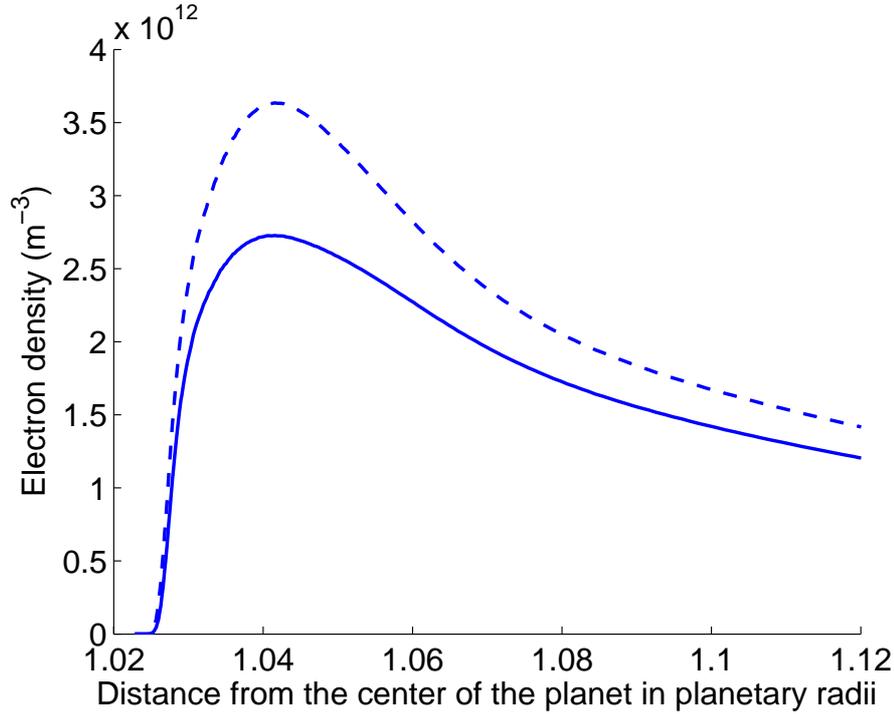}
\caption{Electron density profile for the case of a planet like HD 209458b at 1 AU around a Sun-like star at solar minimum (solid line) and solar maximum (dashed line) for a latitude of $30^\circ$. The standoff distances for $0.1 \mathcal{M}_{\rm J}$ and for the Jovian magnetic moment would be at $\approx 8.3 R_{\rm p}$ and $\approx 17.9 R_{\rm p}$, respectively. The curves end at the exobase level.}
\label{fig:fig8}
\end{center}
\end{figure*}

Figure \ref{fig:fig8} shows the electron densities at 30 degrees latitude on the dayside of the planet calculated with the model of \citet{Chadney2015} for the case of an HD 209458b-like planet at 1 AU around a Sun-like star for solar minimum and maximum. The curves stop above the exobase levels at 1.12 $R_{\rm p}$ while the magnetopause standoff distance would be located at about $\approx 8.3 R_{\rm p}$ for $0.1 \mathcal{M}_{\rm J}$. If HD 209458b had a Jovian magnetic moment, the standoff distance would be at $\approx 17.9 R_{\rm p}$. For Jupiter, standoff distances of up to $100 R_{\rm p}$ have been observed, i.e. there exists a very large magnetospheric cavity with large regions of dilute plasma, in contrast to Hot Jupiters.

\begin{figure*}
\begin{center}
\includegraphics[width=1.00\columnwidth]{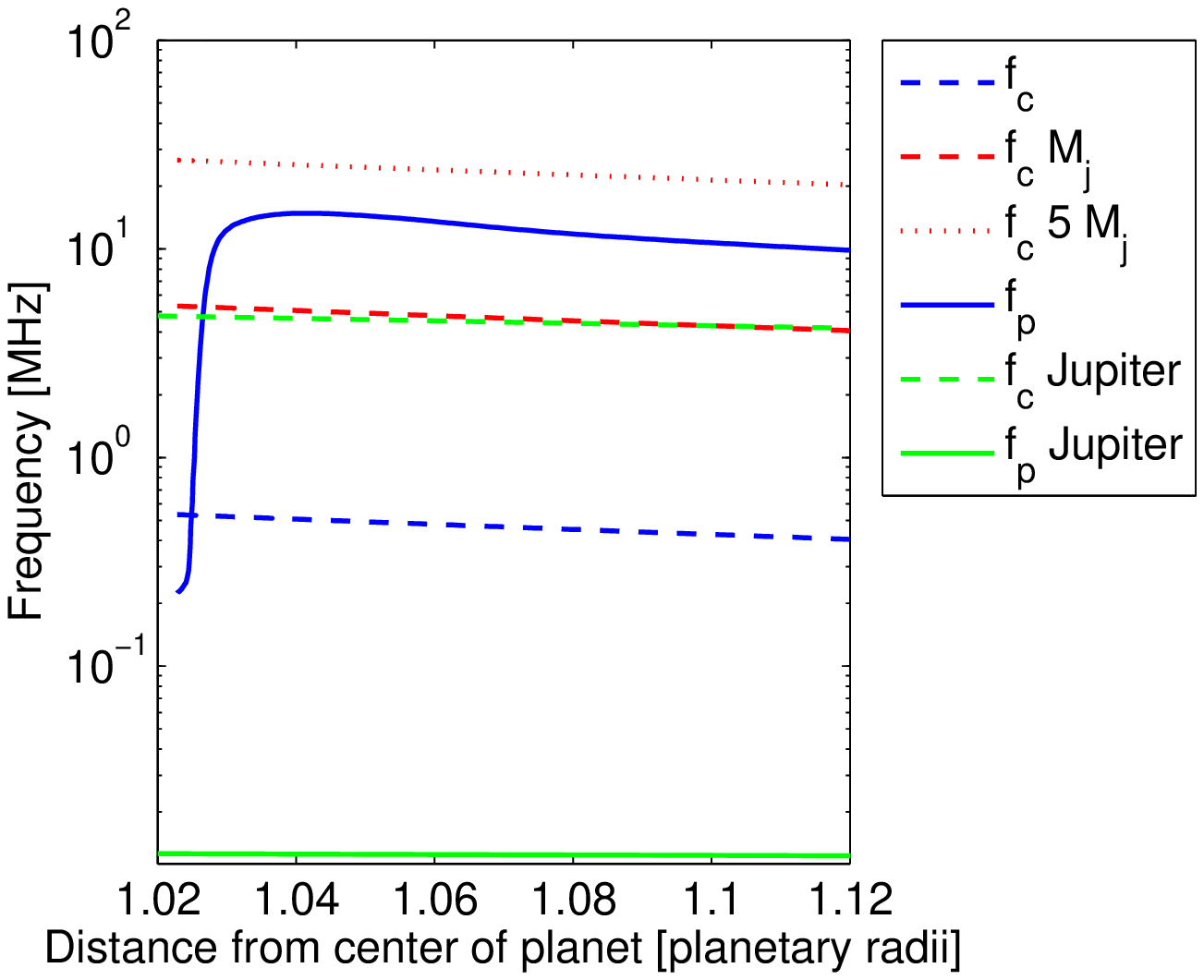}
\includegraphics[width=1.00\columnwidth]{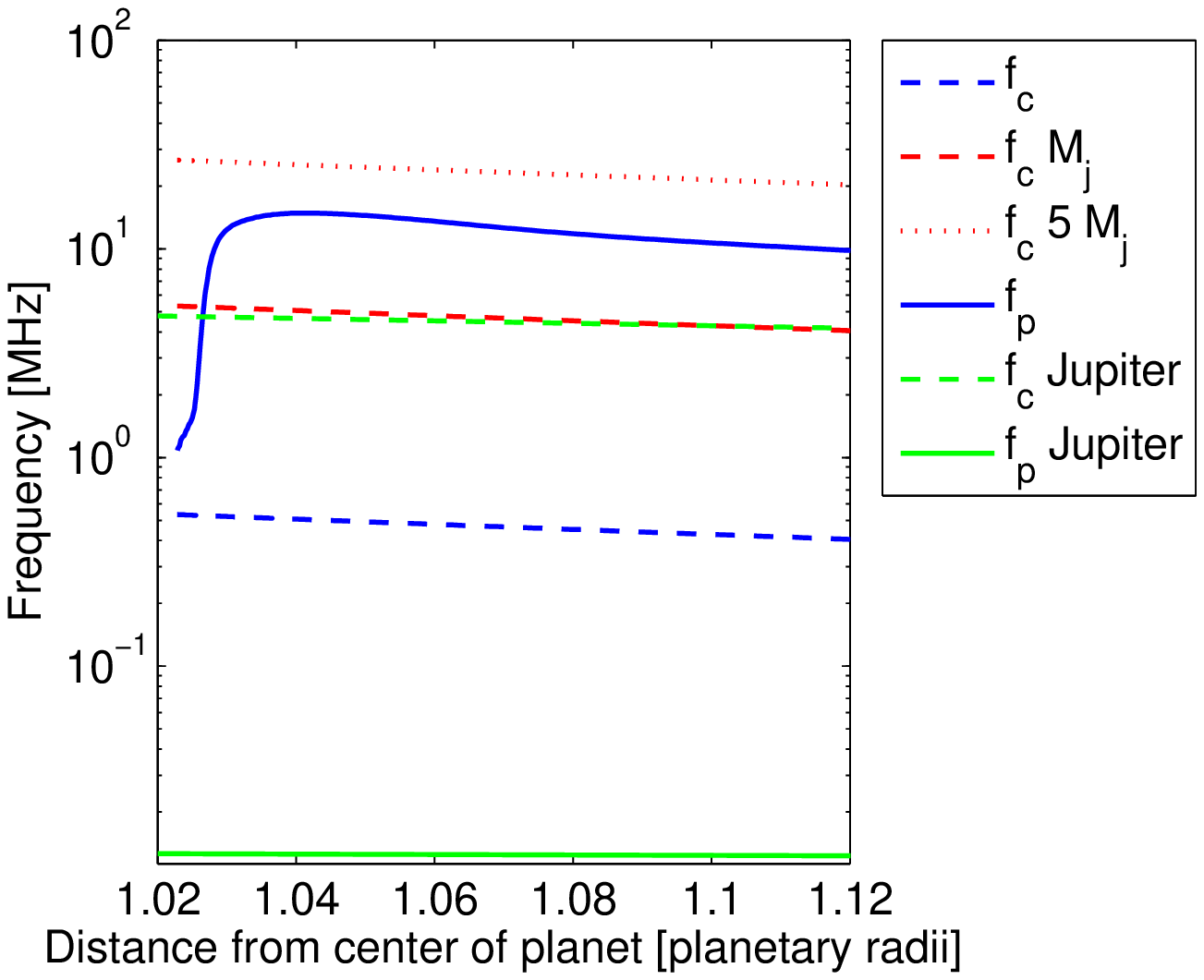}
\caption{Plasma frequency (blue line) and cyclotron frequency (dashed blue, red dashed and red dotted line) for three different magnetic field cases for the 1 AU-case of an HD 209458b-like planet around a Sun-like star for solar minimum on the dayside (left) and nightside (right) of the planet. The dashed blue line corresponds to the magnetic moment of HD 209458b as estimated by \citet{Kislyakova2014}, the red dashed line to the Jovian magnetic moment and the red dotted line to 5 times the Jovian value. The green solid and dashed lines indicate the plasma and cyclotron frequencies at Jupiter, respectively.}
\label{fig:fig9}
\end{center}
\end{figure*}

Figure \ref{fig:fig9} shows the plasma and cyclotron frequencies for the case of solar minimum conditions on the dayside (left) and nightside (right) of the HD 209458b-like planet at 1 AU. The results for the case of solar maximum are not shown because they do not differ significantly. Neither generation nor escape of radio emission are possible for $0.1 \mathcal{M}_{\rm J}$ in the region below the exobase. For $\mathcal{M}_{\rm J}$ generation is possible close to the planet. Only for the case of 5 times the Jovian magnetic moment the cyclotron frequency is larger than the plasma frequency on the day- as well as the night-side of the planet. Thus, generation and escape of radio waves is possible for $5 \mathcal{M}_{\rm J}$. However, the region above the exobase level is not shown due to lack of data. For the case of Jovian magnetic moment and maybe also for $0.1 \mathcal{M}_{\rm J}$ we expect the cyclotron frequency to be higher than the plasma frequency closely above the exobase level. Also the fact that the standoff distance is about 8 times larger than the exobase level for $0.1 \mathcal{M}_{\rm J}$ and at $\approx 17.9 R_{\rm p}$ for the Jovian magnetic moment, i.e. already a large magnetospheric cavity compared to close-in planets, should lead to more favourable conditions for the CMI. At the orbit of 1 AU we already expect similar conditions as for Earth or Jupiter (i.e. very low plasma densities above the exobase level). Thus, the ratio of plasma to cyclotron frequency for the case of Jovian and maybe even for 0.1 times the Jovian magnetic moment is expected to fall below 0.4 closely above the exobase.


At orbit locations greater than 1 AU, one can already expect similar conditions as for Jupiter in the solar system or conditions comparable to Earth. In general one can conclude from our results that exoplanets beyond 0.5 AU around Sun-like stars (for more active stars this value is higher) are more favourable candidates for future radio observations than Hot Jupiters. This result coincides with that of \citet{Nichols2011,Nichols2012} and \citet{Noyola2014,Noyola2015,Noyola2016} who also favour planets at larger orbits (albeit for different reasons).

\section{Discussion and Conclusion}
\label{sec:sec5}

\begin{figure*}
\centering
\includegraphics[width=1.0\columnwidth]{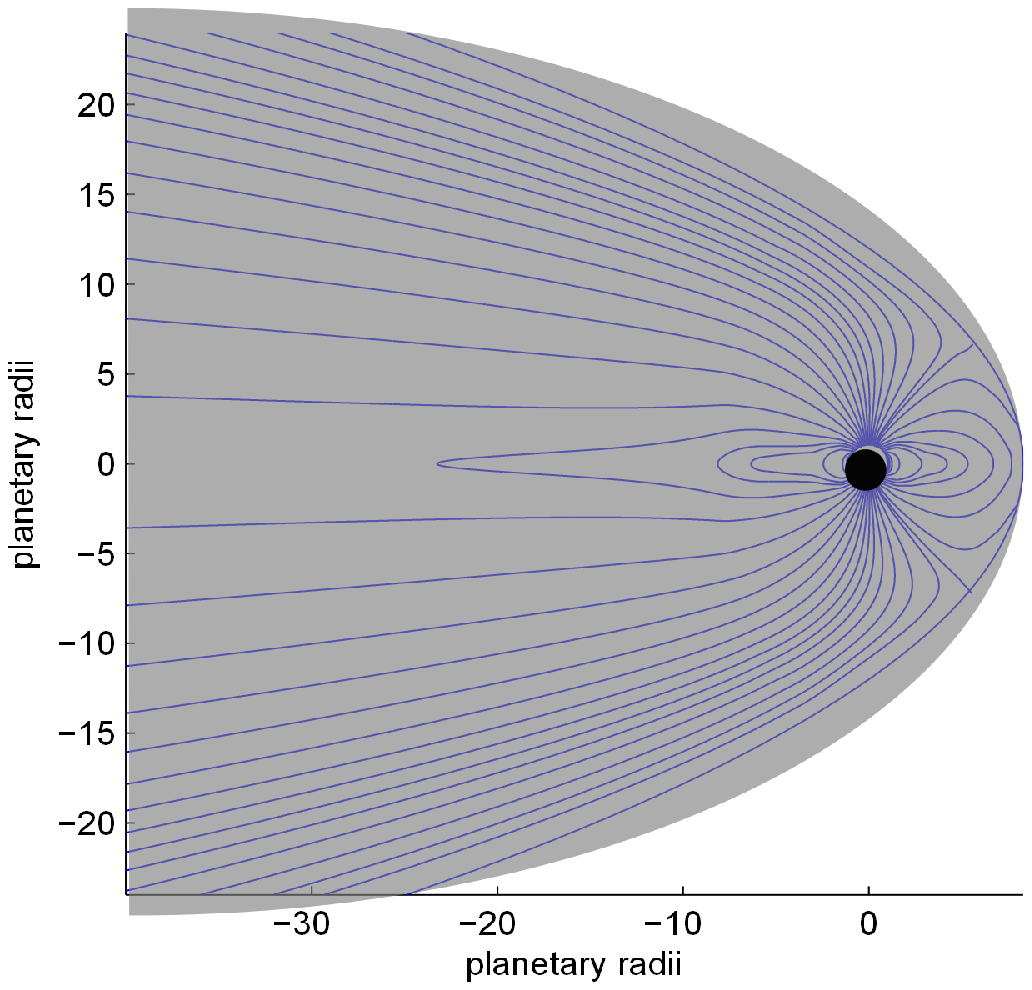}
\includegraphics[width=1.0\columnwidth]{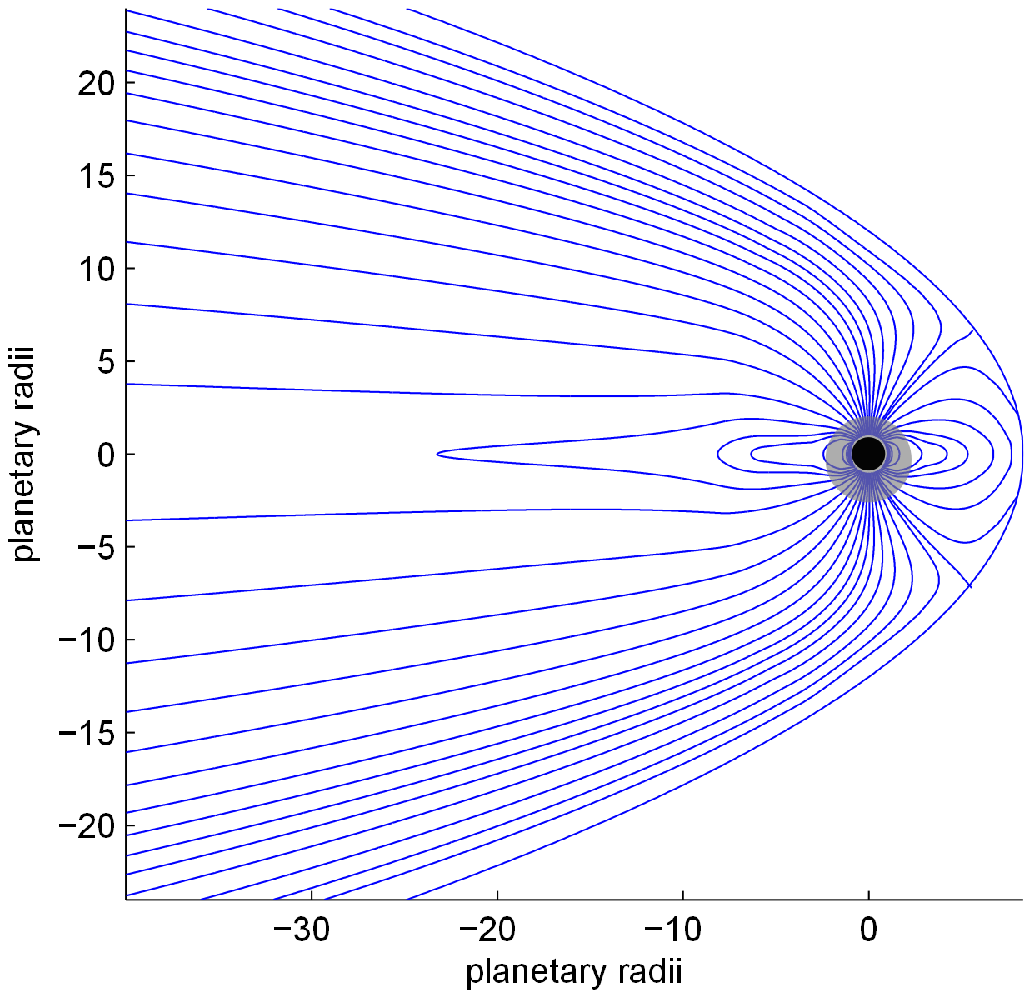}
\caption{Quantitative sketch of the configuration of the magnetosphere for a close-in gas giant (left) and a gas giant at 1 AU (right). The planet is located at the origin of the coordinate system and the axes are scaled in planetary radii. The star is located to the right. Left: grey shaded area indicating the dense plasma filling up the magnetosphere up to its boundary. Right: grey circle indicating the region of dense plasma up to the exobase. Outside of the exobase the plasma has a much lower density.}
\label{fig:fig10}
\end{figure*}

Our results indicate that high ionospheric plasma densities in hydrodynamically expanded upper atmospheres of close-in extrasolar giant planets prevent the radio emission from escaping the source region or render the generation of radio waves via the CMI mechanism impossible at all. Figure \ref{fig:fig10} shows a sketch of the magnetosphere of a Jupiter-like gas giant under the hydrodynamic (left panel) and hydrostatic (right panel) regimes. The magnetospheres of the gas giants are shown with a grey shaded (left) and a grey circular area indicating the regions of dense plasma. For the close-in gas giant the exobase extends out to the magnetospheric boundary whereas for a gas giant beyond 0.5 AU the configuration becomes Earth-like and thus favours radio emission generation via the CMI.

Table \ref{tab:tab2} shows the values of the magnetopause standoff distances for HD 209458b at different orbital distances from a Sun-like host star in comparison with the location of the exobase. For orbital distances greater than or equal to 0.5 AU the exobase is located well inside the magnetosphere whereas for close-in planets at 0.2 AU or less it would be located far beyond the magnetopause. For more active stars the transition from hydrodynamic to Jeans escape occurs at even larger orbital distances between 0.5 to 1 AU.

\begin{table}
\caption{Magnetopause standoff distances $R_{\rm s}$ for a planet such as HD 209458b (with $0.1 \mathcal{M}_{\rm J}$) around a Sun-like star at different orbital locations and locations of the exobase $R_{\rm exo}$.}
\label{tab:tab2}
\begin{tabular}{|p{0.5 cm}|p{1.45 cm}|p{1.45cm}||p{1.45 cm}|p{1.45 cm}|}
& 0.045 AU & 0.2 AU & 0.5 AU & 1 AU \\
\hline
\hline
$R_{\rm s}$ & $2.8 R_{\rm p}$ & $4.8 R_{\rm p}$ & $6.6 R_{\rm p}$ & $8.3 R_{\rm p}$ \\
\hline
$R_{\rm{exo}}$ & $R_{\rm{exo}} > R_{\rm s}$ & $R_{\rm{exo}} > R_{\rm s}$ & $R_{\rm{exo}} < R_{\rm s}$ & $R_{\rm{exo}} < R_{\rm s}$\\
\hline
\hline
\end{tabular}
\end{table}

Hot Jupiters have been widely considered as the best candidates for the observation of exoplanetary radio emission in the literature \citep{Farrell1999,Farrell2004,Zarka2001,Zarka2004,Zarka2007,Lazio2004,Griessmeier2004,Griessmeier2005,Griessmeier2007c,Griessmeier2007b,Stevens2005,Jardine2008,Reiners2010,Vidotto2012,See2015a,See2015b}. However, our results indicate that the CMI, the process which is the most efficient in the generation of radio emissions in planets of the solar system, most likely does not operate at Hot Jupiters. 

Planets at larger orbital distances, around less active, UV-weak stars might be more promising candidates. On the other hand, at larger orbital distances, planets around UV-active stars become better candidates, especially if they have an internal plasma source similar to the Jupiter-Io system. In the latter case a significant amount of radio emission is due to strong field-aligned currents generated by magnetosphere-ionosphere coupling \citep{Nichols2011}. Especially systems which are close to Earth, like Eps Eridani b \citep[which has also been studied by][]{Noyola2014} are good candidates because the flux densities reaching Earth are evidently higher than for systems farther away. The orbit of Eps Eridani b is located at 3.39 AU, i.e. far beyond the aforementioned transition regions.

We are currently performing a follow-up study on more massive Hot Jupiters like Tau Bootis b with larger magnetic moments and on the exact location of the transition region from hydrodynamic to hydrostatic upper atmosphere conditions. Due to the larger mass and thus larger gravity the atmospheres of such planets should be much less extended than for the Hot Jupiters studied in the current paper. In a further follow-up paper we will consider WASP-33b, a Jovian-like planet with about 2 Jovian masses around an UV-weak star. 

\section*{ACKNOWLEDGMENTS}
C. Weber, H. Lammer, and P. Odert acknowledge support from the FWF project P25256-N27 `Characterizing Stellar and Exoplanetary Environments via Modeling of Lyman-$\alpha$ Transit Observations of Hot Jupiters'.
The authors acknowledge also the support by the FWF NFN projects S11606-N16 `Magnetospheres - Magnetospheric Electrodynamics of Exoplanets and S11607-N16 `Particle/Radiative Interactions with Upper Atmospheres of Planetary Bodies Under Extreme Stellar Conditions'. M. L. Khodachenko and I. F. Shaikhislamov acknowledge the support of the FWF projects I2939-N27, P25587-N27 and P25640-N27, Leverhulme Trust grant IN-2014-016, and the grants No. 16-52-14006, No. 14-29-06036 of the Russian Fund for Basic Research. The authors also thank an anonymous referee for valuable comments and suggestions which helped to improve this paper.  

\bibliography{mybibtex}



\bsp
\label{lastpage}
\end{document}